\documentstyle[prb,eqsecnum,aps,epsfig]{revtex}
\begin{document}
\draft


\title{
Electromagnetic absorption of a pinned Wigner crystal at finite temperatures
}

\author{Hangmo Yi and H.A.\ Fertig}

\address{
Department of Physics and Astronomy and Center for Computational Sciences, University of Kentucky, Lexington, Kentucky 40506
}

\date{\today}

\maketitle

\begin{abstract}
We investigate the microwave absorption of a pinned, 
two-dimensional Wigner crystal in a strong 
magnetic field at finite temperatures.  
Using a model of a 
uniform commensurate pinning potential, we 
analyze thermal broadening of the electromagnetic 
absorption resonance.  Surprisingly, we find 
that the pinning resonance peak should remain 
sharp {\em even when the temperature 
is comparable or greater than the peak frequency.}  
This result agrees qualitatively with recent experimental observations 
of the ac conductivity in two-dimensional hole 
systems in a magnetically induced insulating state.  
It is shown, in analogy with Kohn's theorem, that 
the electron-electron interaction does not 
affect the response of a harmonically pinned Wigner crystal 
to a spatially uniform external field at any temperature.  
We thus focus on anharmonicity in the pinning potential 
as a source of broadening.  Using a $1/N$
expansion technique, we show that the broadening is
introduced through the self-energy corrections to the
magnetophonon Green's functions.
\end{abstract}



\section{Introduction}
\label{sec:introduction}

It is now well appreciated that in certain situations, 
the translational symmetry of an electron gas 
becomes broken at low temperatures giving rise 
to an electron solid state 
called the Wigner crystal (WC). \cite{wigner34}  
Among other conditions, low density and low disorder are crucial 
for this novel state to be realized; the inter-electron distance has 
to be large enough for the Coulomb interaction to dominate the quantum 
zero-point fluctuations, and the disorder must be 
weak enough not to destroy the long- (or quasi-long-) range order.  
One of the systems that satisfy 
these conditions best is the two-dimensional (2D) electron gas in the 
GaAs/AlGaAs heterostructure, in which density and disorder can 
be controlled relatively easily.  
Furthermore, if a strong magnetic 
field is applied perpendicular to the 
2D plane, the cyclotron motion 
quenches zero-point fluctuations and 
the low-density condition is even 
more easily achieved. \cite{WCGeneralFertig97inbook}  

Although direct observations of the positional order from 
scattering measurements, for example, have been unattainable, 
there have been considerable efforts to prove the existence of the 
WC through more or less indirect methods such as 
transport measurements 
\cite{jiang90,goldman90,jiang91,santos92,li97,li98,hennigan98} 
and photoluminescence. 
\cite{buhmann91,clark91,kukushkin92,kukushkin94,goldys92}  
One of the most convincing evidences so far 
is that the 2D systems become 
insulating when the filling factor $\nu=nhc/eB$ is 
low: $\lesssim 1/5$ for electron systems \cite{jiang90} 
and $\lesssim 1/3$ for hole systems. \cite{santos92}  This insulating 
behavior is generally accepted as a result of pinning 
of the WC due to impurities.  The impurity potential 
is also supposed to provide a restoring force in the ac 
electromagnetic response of the WC.  If the frequency 
of the external driving force matches the natural 
frequency of the pinning mode, a resonance should occur.  

Recent experiments on 2D hole systems \cite{li97,li98,hennigan98} 
have revealed a resonance structure in the 
ac absorption spectrum in the low 
filling factor regime.  However, some features 
of the spectrum were qualitatively different from the predictions 
of the previous calculations based on charge-density-wave (CDW) 
models; \cite{fukuyama78a,normand92} 
namely, (i) the peak frequency $\omega_p$ 
increased with the magnetic field $B$, and (ii) the quality 
factor $Q=\omega_p/\Delta\omega$ was found to be 
as large as 30 in some 
data. \cite{hennigan98}  The CDW 
calculations had predicted that $\omega_p$ would decrease with $B$ 
and the peak would be much broader ($Q\sim 1$).  

These puzzling experimental findings were later examined 
by the authors of Refs.\ \ref{ref:chitra98} \nocite{chitra98}
and \ref{ref:fertig99}. \nocite{fertig99}  
Based on weak disorder models at $T=0$, 
they found that (i) was 
explained if the disorder potential was allowed to vary within 
a magnetic length $l_0=\sqrt{\hbar c/eB}$.  This short-length-scale 
physics is missing in the previous CDW calculations since 
they assume that the disorder potential can vary only in a length 
scale longer than the lattice periodicity $a$.  (In 
the experiments in Refs.\ \ref{ref:li97}, \ref{ref:li98}, and 
\ref{ref:hennigan98}, $a>l_0$.)  More specifically, 
in Ref.\ \ref{ref:fertig99}, it was suggested that 
the roughness of the GaAs/AlGaAs interface could serve as a source of 
such a short-length-scale disorder potential.  A ``pit'' 
on the interface, which is typically several atoms wide, 
can trap an electron by allowing it to stay closer to the positively 
charged donor plane and gain electrostatic energy.  
In a strong magnetic field, one can project the 
many-body wave function onto the lowest Landau level (LLL).  
Assuming $a\gg l_0$, one may ignore exchange energy 
and describe the many-body ground state as a Hartree-type 
product of single electron wave packets of size $\sim l_0$, 
which are, in the absence of disorder, located at crystal 
lattice sites. \cite{maki83} 
Suppose an electron wave packet is situated 
right on a pit.  Since its size decreases with 
increasing $B$, the probability for the electron 
to stay inside the pit subsequently increases, 
provided {\em the pit size is smaller than $l_0$}.  
Since $\omega_p$ is essentially given by the expectation 
value of the pinning potential strength, 
the above cartoon picture explains how $\omega_p$ grows 
with $B$.  In addition, as described below, the long-range 
Coulomb interaction suppresses low-energy collective 
excitations and leads to a sharp resonance.  

At finite temperatures, however, thermal fluctuations should 
broaden the resonance peak.  
One important observation is that the peaks 
measured in experiments keep getting 
narrower down to the lowest $T$ measured.  This suggests that 
the main source of broadening is thermal fluctuations.  
Interestingly, the experiments also find 
that $\Delta\omega\ll T$ even though $T\gtrsim\omega_p$.  
Thus, how the thermal broadening comes about is a highly 
non-trivial question.  We address the question in this 
paper.  As the thermal broadening is the main issue, 
we will simplify the disorder model and assume that the 
pinning centers themselves form exactly the same lattice 
as the WC.  We also assume that the impurity potential 
has the same strength at all sites. \cite{seeCote92}  Several different 
disorder potentials --- including the uniform commensurate 
model we will use in this work --- have been studied at $T=0$ in a 
previous study, \cite{fertig99} but it was found that the details of 
disorder do not alter the resonance structure 
qualitatively.  We believe the thermal behavior is 
not sensitive to details of disorder, either.  

Our result can be summarized as follows.  
In order to understand low, but finite temperature behavior, 
one needs to investigate the low-energy excitations 
and their interactions.  The longitudinal and the transverse phonon 
modes of the WC mix together in a strong magnetic field and 
two new normal modes, magnetophonons 
and magnetoplasmons, emerge. \cite{bonsall77,cote90,cote91}  
We focus on the magnetophonons because the 
energy of the other mode, magnetoplasmons, 
is essentially given by the cyclotron 
energy $\hbar\omega_c=\hbar eB/mc$, which is a few orders 
of magnitude greater than $T$ and $\omega_p$.  
In a theory that treats displacement of electrons from 
lattice sites harmonically, magnetophonons are independent 
and produces only a delta-function peak in the ac absorption 
spectrum.  The interaction between magnetophonons 
comes about from anharmonicity of the Hamiltonian.  
There are two sources of anharmonicity in our model: 
Coulomb interaction and pinning potential.  
Since the external ac field has a much larger 
wave length ($\gtrsim 1$ m) than the 
size of the system (typically $\sim 10^{-5}$ m), \cite{li97} 
it is safe to treat it as a uniform field that 
couples only to the center-of-mass degree of 
freedom.  If the pinning potential is completely 
harmonic, we find that the anharmonicity 
in the Coulomb interaction 
does not contribute to the absorption spectrum, 
because the interaction depends only on the 
relative position of electrons.  
This is analogous to the 
Kohn's theorem, \cite{kohn61} which states that 
the electron-electron interaction 
does not affect the cyclotron frequency in a 
disorder-free system.  The theorem remains true even in a 
harmonically pinned WC, 
Therefore, anharmonicity of the pinning 
potential is crucial for understanding thermal broadening.  

As will be shown below, magnetophonon modes are very closely 
related to angular momentum excitations.  It 
is therefore natural to exploit analogies to 2D spin lattice systems.  
In fact, written in magnetophonon creation and annihilation 
operators, our model Hamiltonian resembles that of an SU(N) 
Schwinger boson formulation of a 2D quantum 
(anti-)ferromagnet. \cite{schwinger65book,auerbach94book,timm98}  
Further utilizing this analogy, 
we use a $1/N$ expansion technique that 
has proved useful in spin systems.  
Not only does it provide a systematic way of collecting 
important processes in Feynman diagrams, but also 
its lowest-order mean-field (MF) solution already contains terms 
that are higher order in coupling constant in a standard 
diagrammatic many-body calculation.  Importantly, 
the position of absorption resonance peak is mostly 
determined at the MF level.  

The Coulomb interaction and the pinning potential 
affect the spectrum of the magnetophonons significantly.  
Without pinning, the dispersion relation is well known to 
take the form $\varepsilon_{\mathbf q} \propto |{\mathbf q}|^{3/2}$ 
in the long wave length limit, 
calculated both classically \cite{bonsall77} and 
quantum mechanically \cite{fukuyama75,maki83}.  
In the uniform commensurate pinning model, it is modified 
to $\varepsilon_{\mathbf q} \approx V + v_s |{\mathbf q}|$, 
where $V$ represents the strength of the pinning 
potential and $v_s$ is the sound velocity of 
the magnetophonon mode.  
The new spectrum is not only gapped, but also makes 
the density of states (DOS) vanish linearly 
in the low-energy limit.  It turns out 
that the slope of the low energy DOS is so small 
that there are extremely few low-energy excitation modes 
about $\omega_p$.  
As a consequence, we find that the 
thermal fluctuations are substantially 
suppressed and the peak remains sharp.  

This paper is organized as follows.  
The magnetophonon creation and annihilation 
operators are constructed out of displacement 
operators in the LLL in Sec.\ \ref{sec:magnetophonons}.  
We derive the ac conductivity 
in terms of magnetophonon Green's functions 
in Sec.\ \ref{sec:conductivity}.  
We then introduce in Sec.\ \ref{sec:hamiltonian}, 
the Hamiltonian of the WC, and derive 
in Sec.\ \ref{sec:kohnsTheorem}, 
a generalized Kohn's theorem that 
explains why the anharmonicity in the Coulomb 
interaction does not contribute to thermal 
shift or broadening of the absorption 
spectrum in a harmonic pinning model.  
Sec.\ \ref{sec:1_NExpansion} is 
devoted to $1/N$ expansion calculations 
of the Green's functions.  We present the 
results in Sec.\ \ref{sec:result}, 
and finally conclude 
with a discussion in Sec.\ \ref{sec:conclusion}.  
Details of some calculations are given in 
Appendices.

\section{Magnetophonons}
\label{sec:magnetophonons}

Before we write down the Hamiltonian in a second-quantized form, 
it is helpful to see how the Hilbert space is spanned by 
the eigenfunctions of magnetoplasmon and magnetophonon 
number operators.  In the circular 
gauge, ${\mathbf A}=(1/2){\mathbf B}\times{\mathbf r}$, 
a 2D single electron wavefunction 
in the $n$-th Landau level with angular momentum 
quantum number $m$ takes the form
\begin{equation}
\psi_{nm}({\mathbf r}) = \frac{(-\sqrt{2}l_0)^{2n+m}}{\sqrt{2\pi l_0^2n!(n+m)!}} e^\frac{|z|^2}{4l_0^2} \partial_z^n \partial_{z^*}^{n+m} e^{-\frac{|z|^2}{2l_0^2}},
\end{equation}
where $z=x+iy$, $\partial_z=(\partial/\partial x-i\partial/\partial y)/2$,
 $\partial_{z^*}=(\partial/\partial x+i\partial/\partial y)/2$, 
and $l_0=\sqrt{\hbar c/eB}$ is the 
magnetic length.  This wavefunction spans the entire single particle 
Hilbert space.  It is now useful to define ladder operators 
that raise or lower $n$ and $m$.  We define the Landau level 
raising operator
\begin{equation}
a^\dagger = \frac{1}{\sqrt{2}} \left( \frac{z^*}{2l_0} - 2l_0\partial_{z} \right), \label{eq:magnetoplasmon}
\end{equation}
and the angular momentum raising operator
\begin{equation}
b^\dagger = \frac{1}{\sqrt{2}} \left( \frac{z}{2l_0} - 2l_0\partial_{z^*} \right). \label{eq:magnetophonon}
\end{equation}
It is straightforward to show that
\begin{eqnarray}
& & a^\dagger\psi_{nm}\propto\psi_{n+1,m-1}, \quad a\psi_{0m}=0, \nonumber \\
& & b^\dagger\psi_{nm}\propto\psi_{n,m+1}, \quad b\psi_{n,-n}=0, \nonumber \\
& & [a,a^\dagger] = [b,b^\dagger] = 1, \nonumber \\
& & [a,b] = [a^\dagger,b^\dagger] = [a,b^\dagger] = [a^\dagger,b] = 0.
\end{eqnarray}
In what follows, the above ladder operators will be defined at 
each WC lattice site.  Then, a Fourier transformed 
operator $a^\dagger_{\mathbf q}$ will serve 
as a magnetoplasmon creation operator, 
and $b^\dagger_{\mathbf q}$ as a magnetophonon 
creation operator.  We will use the 
latter terms to refer to the $a^\dagger$ 
and $b^\dagger$ operators from this point on.  

In the low filling factor 
limit, the Landau gap $\hbar\omega_c$ is large, 
so one can safely truncate the Hilbert space and 
work in the LLL only.  Furthermore, the ground state of the 
WC in this limit is believed to be well represented by 
a product of a collection of zero angular momentum 
single particle 
wave functions, each of which is centered at a lattice 
site. \cite{maki83}  This is analogous to a Hartree 
wave function.  Although the single particle wave functions are not 
orthogonal, their overlap is exponentially small if $l_0\ll a$.  
Thus, one may ignore exchange energy correction 
and the wave functions do not 
have to be antisymmetrized.  In other words, electrons are 
distinguishable because each one of them is strongly 
localized at a lattice site far from the others.  
For higher angular momentum states, 
the size of wave packets grow with $m$.  
However, the above argument should be still 
valid for the first few higher angular momentum states.  
Therefore, we assume that the Hilbert space 
is given by a direct product of 
single particle Hilbert spaces spanned by 
angular momentum states in the LLL.  

Due to the strong Coulomb repulsion, we assume that 
there is only one electron per site.  
This allows us to label each many-body energy eigenstate by 
a collection of $N_s$ angular momentum quantum 
numbers or magnetophonon occupation numbers, 
where $N_s$ is the total number of 
sites.  As will be shown below, the above 
statement is not exactly true, because 
in reality, different angular momentum states 
mix together due to quantum fluctuations.  
This is demonstrated in our Hamiltonian 
as terms that do not 
conserve the magnetophonon number.  
Nonetheless, higher angular momentum mixing is small 
at low temperatures and the above argument 
should be a good qualitative description of the system.  

The many-body wavefunction is written as \cite{maki83} 
\begin{equation}
\Psi(\{{\mathbf r}_i\}) = \prod_i \psi_{m_i}({\mathbf r}_i;{\mathbf R}_i),
\end{equation}
where the single particle 
wavefunction $\psi_{m_i}({\mathbf r}_i;{\mathbf R}_i)$ is localized 
at a lattice site ${\mathbf R}_i$ and has an 
angular momentum quantum number $m_i$.  
More specifically, it may be written as
\begin{equation}
\psi_m({\mathbf r}_i;{\mathbf R}_i) = \frac{1}{\sqrt{2\pi l_0^2m!}} \left( \frac{z_i-Z_i}{\sqrt{2}l_o} \right)^m e^{-\frac{|z_i-Z_i|^2}{4l_0^2}} e^{\frac{z_iZ_i^*-z_i^*Z_i}{2l_0^2}}, \nonumber
\end{equation}
where $Z_i=R^x_i+iR^y_i$.  Note that the last exponential is 
a pure phase and is needed to keep the wavefunction in the 
LLL. We will assume that $\{{\mathbf R}_i\}$ forms a triangular 
lattice in the ground state.  The ladder operators in 
Eqs.\ (\ref{eq:magnetoplasmon}) and (\ref{eq:magnetophonon}) 
are modified at each lattice site 
and may be written as
\begin{eqnarray}
a_i^\dagger & = & \frac{1}{\sqrt{2}} \left( \frac{\zeta_i^*+Z_i^*}{2l_0} - 2l_0\partial_{\zeta_i} \right), \label{eq:ladderA} \\
b_i^\dagger & = & \frac{1}{\sqrt{2}} \left( \frac{\zeta_i-Z_i}{2l_0} - 2l_0\partial_{\zeta_i^*} \right), \label{eq:ladderB}
\end{eqnarray}
where $\zeta_i=z_i-Z_i$ is the complex notation 
of the displacement of the $i$th electron from its 
site.  

In terms of the above ladder operators, the 
original electron position is written as
\begin{eqnarray}
x_i & = & R^x_i + \frac{l_0}{\sqrt{2}} (b_i + b_i^\dagger + a_i + a_i^\dagger), \\
y_i & = & R^y_i + \frac{l_0}{\sqrt{2}} (-b_i + b_i^\dagger + a_i - a_i^\dagger).
\end{eqnarray}
When an operator is applied only to 
states that are completely in the LLL, it may be 
represented by a ``LLL-projected'' operator.  In the above 
expression, this means simply dropping 
the magnetoplasmon operators $a_i$ and $a_i^\dagger$ 
that couples the LLL states to higher Landau levels.  
The projected operators then take the form
\begin{eqnarray}
x_i & = & R^x_i + \frac{l_0}{\sqrt{2}} (b_i + b_i^\dagger), \label{eq:xFromB} \\
y_i & = & R^y_i + \frac{l_0}{\sqrt{2}} (-b_i + b_i^\dagger). \label{eq:yFromB}
\end{eqnarray}
The above operators satisfy the usual 
LLL commutation relation \cite{conti98}
\begin{equation}
[x_i,y_j] = l_0^2 \delta_{ij}. \label{eq:commutationXY}
\end{equation}

\section{Conductivity}
\label{sec:conductivity}

The quantity we will eventually compute is the ac 
conductivity tensor $\sigma^{\mu\nu}(\omega)$, 
where $\mu,\nu=x,y$.  In this section, we 
will show how $\sigma^{\mu\nu}(\omega)$ is derived 
in terms of the magnetophonon Green's functions.  

The external perturbation due to a 
spatially uniform electric field with frequency $\omega$ 
is $H'(t)=e{\mathrm Re}\,e^{-i\omega t} {\mathbf E}_0\cdot\sum_i{\mathbf u}_i$,  
where ${\mathbf u}_i={\mathbf r}_i-{\mathbf R}_i$.  
In the linear response theory, 
the electric current is given by
\begin{eqnarray}
j^\mu(t) & = & {\mathrm Re}\,\frac{i\omega e\rho_0}{N_s} \sum_i \langle u^\mu_i(t) \rangle \\
& = & {\mathrm Re}\, \frac{i\omega e\rho_0}{N_s} \sum_i \left(-\frac{i}{\hbar}\right) \int dt' \theta(t-t') \langle [u^\mu_i(t),H'(t')] \rangle \\
& = & {\mathrm Re}\, \frac{i\omega e^2\rho_0}{\hbar} e^{-i\omega t} \chi_R^{\mu\nu}(\omega) E_0^\nu, \label{eq:current}
\end{eqnarray}
where $\rho_0$ is the 2D number density of 
electrons.  The brackets $\langle\cdots\rangle$ denote a 
thermal average for the unperturbed Hamiltonian 
and the zero wave vector response function 
is defined as
\begin{equation}
\chi_R^{\mu\nu}(\omega) = \int dt' e^{i\omega(t-t')} \theta(t-t') \langle [u^\mu_{\mathbf q=0}(t), u^\nu_{\mathbf q=0}(t')] \rangle. \label{eq:chiR1}
\end{equation}
The linear ac conductivity 
can be read off from Eq.\ (\ref{eq:current}) as
\begin{equation}
\sigma^{\mu\nu}(\omega) = -{\mathrm Im}\,\frac{\omega e^2\rho_0}{\hbar} \chi_R^{\mu\nu}(\omega). \label{eq:sigma}
\end{equation}
In general, one has to include the 
polarization field in the calculation and 
solve for the conductivity self-consistently.  
In a system with uniform background charge, 
however, the ${\mathbf q=0}$ component of the 
polarization is meaningful only on the boundaries.  
In 2D, the boundary effect dies off inversely with the size of 
the system, and may be ignored in the thermodynamic 
limit.  

The response function $\chi_R^{\mu\nu}$ 
may be most easily obtained by computing 
the Matsubara frequency response function
\begin{equation}
\chi^{\mu\nu}(i\omega_n) = -i\int dt' e^{i\omega_n\tau} \langle T_\tau u^\mu_{\mathbf q=0}(\tau) u^\nu_{\mathbf q=0}(0) \rangle
\end{equation}
and then analytically continuing
\begin{equation}
\chi_R^{\mu\nu}(\omega) = \chi^{\mu\nu}(i\omega_n\rightarrow\omega+i0^+), \label{eq:chiR}
\end{equation}
where $T_\tau$ denotes time-ordering 
in the imaginary time $\tau$.  
Since $u^\mu_i=r^\mu_i-R^\mu_i$ is 
a linear combination of $b_i$ and $b_i^\dagger$ 
as in Eqs.\ (\ref{eq:xFromB}) and (\ref{eq:yFromB}), $\chi^{\mu\nu}(i\omega_n)$ 
can be written in terms of magnetophonon 
Green's functions as
\begin{eqnarray}
\chi^{\mu\nu} & = & \left[ \begin{array}{cc} \chi^{xx} & \chi^{xy} \\ \chi^{yx} & \chi^{yy} \end{array} \right] \nonumber \\
& = & \frac{l_0^2}{2} \left[ \begin{array}{cc} 1 & 1 \\ i & -i \end{array} \right] \left[ \begin{array}{cc} G^{-+} & G^{--} \\ G^{++} & G^{+-} \end{array} \right] \left[ \begin{array}{cc} 1 & -i \\ 1 & i \end{array} \right]. \label{eq:chi}
\end{eqnarray}
The Green's functions are defined as
\begin{equation}
\left[ \begin{array}{cc} G^{-+} & G^{--} \\ G^{++} & G^{+-} \end{array} \right] = -\int d\tau e^{i\omega_n\tau} \left\langle T_\tau \left[ \begin{array}{cc} b b^\dagger & b b \\ b^\dagger b^\dagger & b^\dagger b \end{array} \right] \right\rangle, \label{eq:green}
\end{equation}
where $b b^\dagger$ 
should be interpreted as $b(\tau) b^\dagger(0)$, etc.

\section{Hamiltonian}
\label{sec:hamiltonian}

The discussions so far have been quite general 
and independent of the actual Hamiltonian 
of the system.  In this section, we will construct 
the Hamiltonian in terms of magnetophonon operators.  
Let us first consider a WC without disorder.  
The kinetic part of the Hamiltonian is given by
\begin{eqnarray}
H_k & = & \frac{\hbar^2}{2m} \left( \nabla + \frac{e}{c}{\mathbf A} \right)^2 \\
& = & \hbar\omega_c \left( a_i^\dagger a_i+\frac{1}{2} \right),
\end{eqnarray}
where $a_i^\dagger$ is 
the magnetoplasmon creation operator 
in Eq.\ (\ref{eq:ladderA}).  
The Coulomb interaction takes the form
\begin{equation}
H_C = \frac{1}{2} \sum_{i\neq j} \frac{e^2}{\kappa|{\mathbf r}_i-{\mathbf r}_j|}, \label{eq:HC0}
\end{equation}
where $\kappa$ is the dielectric constant.  
Expanded in terms of the 
displacement ${\mathbf u}_i$, 
the above equation may be written as
\begin{equation}
H_C = \frac{1}{2} \sum_{i\neq j} \sum_{\mu\nu} u_i^\mu P_{ij}^{\mu\nu} u_j^\nu + \text{const.} + {\mathcal O}\left[ \left( \frac{u}{a} \right)^3 \right], \label{eq:HC}
\end{equation}
where $\mu,\nu=x,y$, and $a$ is the lattice constant.  
The dynamic matrix is given by
\begin{equation}
P_{ij}^{\mu\nu} = \frac{e^2}{\kappa|{\mathbf R}_{ij}|^3} \left( \delta_{\mu\nu} - 3\frac{R_{ij}^\mu R_{ij}^\nu}{|{\mathbf R}_{ij}|^2} \right), \label{eq:dynamicMatrix}
\end{equation}
where ${\mathbf R}_{ij}\equiv{\mathbf R}_i-{\mathbf R}_j$.  

Now we take the strong-magnetic-field limit and 
project the above Hamiltonian to the LLL.  
Technically, this means rewriting it in terms of 
ladder operators in Eqs.\ (\ref{eq:ladderA}) and (\ref{eq:ladderB}),
and abandoning any term that is normal-ordered in $a^\dagger$ 
and $a$, i.e., any term in which all $a$'s are placed to the right 
of all $a^\dagger$'s.  Obviously, the kinetic term is constant 
after the projection, and may be ignored.  The Coulomb 
interaction may be rewritten as
\begin{eqnarray}
H_{C2} & = & - \sum_{i\neq j} \frac{e^2l_0^2}{4|{\mathbf R}_{ij}|^3} \left[ b_i^\dagger b_j + b_i b_j^\dagger + 3\left( n_{ij}^2 b_i b_j + n_{ij}^{*2} b_i^\dagger b_j^\dagger \right) \right] \nonumber \\
& &  + \sum_i \left( \sum_{j(\neq i)} \frac{e^2l_0^2}{2|{\mathbf R}_{ij}|^3} \right) b_i^\dagger b_i, \label{eq:HC2}
\end{eqnarray}
where $n_{ij} \equiv (R_{ij}^x + iR_{ij}^y)/|{\mathbf R}_{ij}|$.  
The above expression may also be derived by making use 
of the analogy between the electron system and 
a spin-lattice system (see Ref.\ \ref{ref:fertig99}).  
Note that rewriting Eq.\ (\ref{eq:HC}) into Eq.\ (\ref{eq:HC2}), 
we have dropped not only a constant term, but also the last term that 
is anharmonic.  In fact, in order to investigate 
thermal broadening, it is necessary 
to keep at least one anharmonic term in the Hamiltonian, 
because a completely harmonic Hamiltonian would lead to a theory 
of free independent eigenmodes and all correlation 
functions would be temperature independent.  
However, the anharmonic terms in the Coulomb interaction 
do {\em not} contribute to thermal shift or broadening 
of the response function to a spatially uniform external 
perturbation in a harmonic pinning 
model.  This is a highly non-trivial statement, 
which we discuss in the next section.  
As a consequence of this, 
we are led to investigate anharmonicity in the other term 
in the Hamiltonian: the pinning potential.  

For simplicity, we 
consider only the uniform commensurate pinning 
model in this work.  In reality, of course, 
disorder is more complicated in several ways.  
First, non-commensurate or non-uniform disorder 
will break the discrete translational symmetry and 
deform the lattice.  Unless the impurity potential 
is exactly circular at each site, it will also mix different 
angular momentum states.  

However, it was shown in Ref.\ \ref{ref:fertig99} that 
at $T=0$, the uniform commensurate pinning model yields qualitatively 
the same result as more complicated disorder models.  
Furthermore, experimental data show that 
the resonance peaks keep getting narrower down 
to $T\lesssim \hbar\omega_p$, suggesting that thermal fluctuation 
affects broadening more than disorder does, 
at least at the observed temperatures.  Therefore, 
we believe this simple disorder model is sufficient for our 
purpose.  

\begin{figure}
\epsfxsize=3in
\centerline{ \epsffile{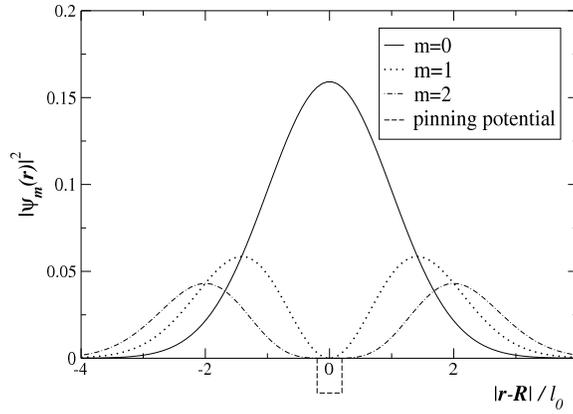} }
\caption{
Probability $|\psi_m({\mathbf r})|^2$ of single particle 
wave functions in the lowest Landau level.  A pit is also 
schematically drawn (a square well at the origin) 
in order to emphasize that 
we are considering the case in which the pit 
size $<l_0$.  States with $m>0$ have little 
overlap with the pit.}
\label{fig:pinning}
\end{figure}

Fig.\ \ref{fig:pinning} shows a schematic picture 
of the probability function for the three 
lowest angular momentum 
states centered at an interface pit.  
Since states with $m>0$ have little probability 
to reside in the pit, they are not affected by it.  
The pinning Hamiltonian is thus appropriately described 
in terms of magnetophonon operators by
\begin{equation}
H_p = -V \sum_i \delta_{b_i^\dagger b_i,0},
\end{equation}
where $V$ is the amount of energy an electron 
pays to move from $m=0$ state to a higher 
angular momentum state {\em in the absence of 
Coulomb interaction}.  This potential is 
highly anharmonic, and the resulting phonon 
problem is exceedingly difficult to 
analyze.  To overcome this, we adopt 
an approximate form of the pinning potential.  

The simplest choice that includes anharmonicity is
\begin{equation}
H_{p4} = -\frac{V}{2} \sum_i (1-b_i^\dagger b_i) (2-b_i^\dagger b_i), \label{eq:Hp4}
\end{equation}
which takes the same value 
as $H_p$ at $b_i^\dagger b_i=0$, 1, and 2.  
At first, the above Hamiltonian 
looks problematic because it has 
no lower bound.  As will be shown below, 
however, the $1/N$ expansion is well defined 
provided the thermal 
average $\langle b_i^\dagger b_i \rangle < 1$.  
It turns out that $\langle b_i^\dagger b_i \rangle$ 
is indeed small and this approximation gives 
sensible results in the observed temperature 
range $T\lesssim 10\hbar\omega_p$.  

Finally, the total Hamiltonian of our model 
is given by
\begin{equation}
H = H_{C2} + H_{p4}. \label{eq:HTotal}
\end{equation}
A similar Hamiltonian was derived in 
Ref.\ \ref{ref:fertig99} by limiting the Hilbert 
space to two angular momentum states per site 
and writing the Hamiltonian in a pseudospin representation.  
Although the higher-angular-momentum states 
should not affect low temperature behavior qualitatively, 
and the assumption that the many-body wave function 
takes a simple product form breaks down for large $m_i$, 
the above Hamiltonian has been derived 
without restricting the Hilbert space to a fixed number 
of states per site.

\section{Generalized Kohn's theorem}
\label{sec:kohnsTheorem}

In this section, we will derive a generalized 
Kohn's theorem, which guarantees that the anharmonicity 
in the Coulomb interaction does not alter the 
absorption spectrum of a uniform electromagnetic 
field for a harmonically and uniformly 
pinned model.  Similar generalizations have been 
made in the context of an electron gas in 
a parabolic potential in Refs.\ \ref{ref:brey89} \nocite{brey89} 
and \ref{ref:yip91}. \nocite{yip91}  

Before we present the proof, let us 
take a brief look at the original Kohn's theorem.  
In a magnetic field, the energy spectrum of a 
single free electron is simply given by Landau levels.  
Now, suppose that there are many electrons 
interacting with one another via Coulomb or 
any other interaction.  
Kohn's theorem tells us that the electromagnetic 
absorption spectrum of such a system, 
as measured through a response to a spatially uniform 
external field, will be exactly 
the same as that of a single electron.  The 
proof of this theorem is disarmingly simple.  
Since the electron-electron interaction depends only on 
the relative degrees of freedom, it commutes 
with the center-of-mass motion.  
Therefore, the observed center-of-mass spectrum 
is not affected by the interaction at all.  

The big difference in the problem considered in this 
paper is that there is pinning.  However, a similar 
argument can be established in a special case 
of harmonic uniform pinning model.  Suppose 
that the pinning Hamiltonian takes the form
\begin{eqnarray}
H_{\mathrm har} & = & V \sum_i b_i^\dagger b_i \\
& = & V \sum_{\mathbf q} b_{\mathbf q}^\dagger b_{\mathbf q}.
\end{eqnarray}
Since this Hamiltonian is separable in 
different momenta $\mathbf q$, its 
eigenstates may be written as a product of 
the form
\begin{equation}
|\Psi\rangle = |\Psi_{\mathrm CM}\rangle |\Psi_{\mathrm rel}\rangle, \label{eq:eigenstates}
\end{equation}
where $|\Psi_{\mathrm CM}\rangle$ is an eigenstate 
of the center-of-mass 
operator $b_{\mathbf q=0}^\dagger b_{\mathbf q=0}$ 
and $|\Psi_{\mathrm rel}\rangle$ of 
all the other relative degrees of freedom.  

Suppose that we calculate the ac conductivity of 
a {\em non-interacting} electron system using 
the expressions in Sec.\ \ref{sec:conductivity}.  
In order to compute the thermal average, one 
needs to compute the expectation values over 
the above eigenstates.  
We make an important observation here that the conductivity 
depends only on $|\Psi_{\mathrm CM}\rangle$, 
because the operators in Eq.\ (\ref{eq:chiR1}) 
are functions only of $b_{\mathbf q=0}^\dagger$ 
and $b_{\mathbf q=0}$.  

Now let us add the Coulomb interaction as in 
Eq.\ (\ref{eq:HC0}) to the Hamiltonian.  
As $H_C$ depends only on relative degrees of 
freedom, it obviously commutes with the 
center-of-mass displacement
\begin{equation}
{\mathbf u}_{\mathrm CM} = \frac{1}{N_s} \sum_i {\mathbf u}_i.
\end{equation}
In terms of magnetophonon operators, it implies
\begin{equation}
[H_C, b_{\mathbf q=0}^\dagger b_{\mathbf q=0}] = 0.
\end{equation}
Therefore, the Coulomb interaction does not alter 
the spectrum of $|\Psi_{\mathrm CM}\rangle$, although 
it may (and in general, it does) 
alter that of $|\Psi_{\mathrm rel}\rangle$.  
Since the ac conductivity is determined 
by the spectrum of $|\Psi_{\mathrm CM}\rangle$ 
alone, it is unaltered 
by the presence of the Coulomb interaction.  
This completes the 
proof of the generalized Kohn's theorem.  

This theorem implies that if we use a 
harmonic pinning potential, the contribution 
from the anharmonic Coulomb interaction terms 
in Eq.\ (\ref{eq:HC}) to the uniform response 
function will {\em vanish in all orders} of $u/a$.  
Therefore, we will use a harmonic Hamiltonian for 
Coulomb interaction, but an anharmonic one for 
pinning potential, hence the total 
Hamiltonian is given as in Eq.\ (\ref{eq:HTotal}).  
The detailed analysis of the anharmonic pinning 
potential will be given below, using a $1/N$ expansion 
technique.

\section{$1/N$ expansion}
\label{sec:1_NExpansion}

A $1/N$ expansion technique consists of 
three steps.  First, we make $N$ copies 
of the magnetophonon operators.  Second, using 
a Hubbard-Stratonovich (HS) decoupling technique, the anharmonic 
terms are replaced by products of a harmonic term 
and an auxiliary HS field.  Finally, 
we expand the magnetophonon Green's function 
in powers of $1/N$.  In principle, the expansion 
is exact if all orders of $1/N$ are kept.  Instead, 
as is conventionally done, we will stop at the first order 
to get an approximation.  Before going into details, 
it should be noted that $N$ will be set equal to 1 
at the end of the approximation, therefore it is 
not a perturbative expansion in any sense.  However, 
this technique usually does a better job in systematically 
picking out important processes than many other many-body 
techniques.  In fact, the zeroth order approximation, which is 
a harmonic theory with magnetophonons coupled to the 
MF value of the HS 
field, already contains the RPA contribution.  The 
first order approximation adds self-energy corrections 
to the Green's functions and gives us 
a thermal broadening.  

After replication, Eq.\ (\ref{eq:HTotal}) becomes
\begin{eqnarray}
H & = & \sum_{l=1}^N H_{C2}(\{b_{li}, b_{li}^\dagger\}) \nonumber \\
& & \quad + V \sum_{li} b_{li}^\dagger b_{li} - \frac{V}{2N} \sum_{l_1l_2i} b_{l_1i}^\dagger b_{l_2i}^\dagger b_{l_2i} b_{l_1i},
\end{eqnarray}
where $l$, $l_1$, and $l_2$ are species indices.  $H_{C2}$ 
is the same as in Eq.\ (\ref{eq:HC2}) 
with different magnetophonon operators 
substituted for each species.  There 
is an extra factor of $1/N$ in the last term 
because it involves two species summations.  
Note that the above equation has been normal-ordered.  
In this form, we may recast the 
problem in a path-integral form simply 
by replacing $b$ ($b^\dagger$) by a classical 
field $z(\tau)$ ($z^*(\tau)$)
in the imaginary time $\tau=it$.  The Euclidean 
action is given by
\begin{eqnarray}
S & = & \int d\tau \Biggl\{ \sum_{li} z_{li}^*(\tau)\partial_\tau z_{li}(\tau) + \sum_l H_{C2}(\{z_{li}(\tau),z^*_{li}(\tau)\}) \nonumber \\
& & + V \sum_{li} |z_{li}(\tau)|^2 - \frac{V}{2N} \sum_i \left[ \sum_l |z_{li}(\tau)|^2 \right]^2 \Biggr\}.
\end{eqnarray}
The first term is the usual Berry phase in the 
imaginary time.  

In the second step, we introduce a real HS 
field $Q_i(\tau)$ to decouple the quartic 
term.  The procedure is formally expressed 
in a functional integral form as
\begin{eqnarray}
\exp & \displaystyle \Biggl[ & \int d\tau \frac{V}{2N} \biggl( \sum_l |z_{li}|^2 \biggr)^2 \Biggr] \nonumber \\
& & = c \int {\mathcal D}Q_i\, \exp \Biggl[ -\int d\tau \nonumber \\
& & \qquad\qquad\qquad\qquad \frac{V}{2} \biggl( -2Q_i\sum_l |z_{li}|^2 + NQ_i^2 \biggr) \Biggr], \nonumber
\end{eqnarray}
where $c$ is a constant.  
Then the generating functional written in 
terms of both $z$ and $Q$ takes the form
\begin{equation}
Z \equiv \int {\mathcal D}^2z {\mathcal D}Q\, e^{-S_{\mathrm HS}},
\end{equation}
where
\begin{eqnarray}
S_{\mathrm HS} & = & \int d\tau \Biggl[ \sum_{li} z_{li}^*\partial_\tau z_{li} + \sum_l H_{C2}(\{z_{li},z^*_{li}\}) \nonumber \\
& & \quad + V \sum_{li} (1-Q_i) |z_{li}|^2 + \sum_i \frac{NV}{2} Q_i^2 \Biggr]. \label{eq:SHS}
\end{eqnarray}
${\mathcal D}^2z$ denotes functional 
integral over real and imaginary parts 
of $z$ field.  So far the action contains all orders 
of $1/N$ and is therefore exact.  
In the following subsections, we will 
Taylor-series expand it in powers of $1/N$.

\subsection{Mean field theory}
\label{subsec:MFTheory}

As is the usual practice, \cite{auerbach94book} 
we assume that the MF solution of $Q_i(\tau)$ 
is independent of $i$ and $\tau$ 
(Ref.\ \ref{ref:meanField}). \nocite{meanField}  
Since the crystal momentum ${\mathbf q}$ is still 
a good quantum number, it is useful 
to Fourier-transform the action.  
We define
\begin{equation}
z_{lq} = \frac{1}{\beta\sqrt{N_s}} \int d\tau \sum_i e^{i\omega_n\tau-i{\mathbf q}\cdot{\mathbf R}_i} z_{li},
\end{equation}
where $q$ is a shorthand notation 
for $(i\omega_n,{\mathbf q})$ and $\beta$ 
is the inverse temperature.  
In terms of the above Fourier components, 
the MF action is given by
\begin{eqnarray}
S_{\mathrm MF}(\overline{Q}) & = & \beta \sum_l \sum_q \Bigl[ \bigl( -i\omega_n + f_{\mathbf q} \bigr) z_{lq}^* z_{lq} \nonumber \\
& & \quad\quad - \frac{1}{2} \bigl( g_{\mathbf q} z_{l,-q} z_{lq} + g^*_{\mathbf q} z_{lq}^* z_{l,-q}^* \bigr) \Bigr], \label{eq:SMF}
\end{eqnarray}
where
\begin{eqnarray}
f_{\mathbf q} & = & (1-\overline{Q})V + \frac{1}{2} \sum_{{\mathbf R}_j\neq 0} \left( 1 - e^{-i{\mathbf q}\cdot{\mathbf R}_j} \right) \frac{e^2l_0^2}{|{\mathbf R}_{ij}|^3}, \\
g_{\mathbf q} & = & \frac{3}{2} \sum_{{\mathbf R}_j\neq 0} e^{-i{\mathbf q}\cdot{\mathbf R}_j} \frac{e^2l_0^2}{|{\mathbf R}_{ij}|^3} n_i^2.
\end{eqnarray}
The MF Green's functions take the form
\begin{eqnarray}
& & \left[ \begin{array}{cc} G^{-+} & G^{--} \\ G^{++} & G^{+-} \end{array} \right] = \nonumber \\
& & \qquad \frac{1}{\omega_n^2+f_{\mathbf q}^2-|g_{\mathbf q}|^2} \left[ 
  \begin{array}{cc}
  -i\omega_n-f_{\mathbf q} & g_{\mathbf q}^* \\
  g_{\mathbf q} & i\omega_n-f_{\mathbf q}
  \end{array}
\right]. \label{eq:MFGreensFunc}
\end{eqnarray}

We choose the MF value $\overline{Q}$ in 
such a way that the MF free energy
\begin{equation}
F_{\mathrm MF}(\overline{Q}) = -\ln \int {\mathcal D}^2z\, e^{-S_{\mathrm MF}(\overline{Q})}
\end{equation}
has a saddle point.  
Solving $\partial F_{\mathrm MF}/\partial\overline{Q}=0$, 
we get
\begin{equation}
\overline{Q} = \frac{1}{N_s} \sum_{\mathbf q} \frac{\int {\mathcal D}^2z\, e^{-S_{\mathrm MF}(\overline{Q})} z_{\mathbf q}^* z_{\mathbf q}}{\int {\mathcal D}^2z\, e^{-S_{\mathrm MF}(\overline{Q})}}.
\end{equation}
Note that $\overline{Q}$ above is equal to 
the thermally averaged value of the number of 
magnetophonons per 
site, $n_{\mathrm ph} \equiv \sum_{\mathbf q} \langle b_{\mathbf q}^\dagger b_{\mathbf q} \rangle / N_s$.  
$\overline{Q}$ has to be determined 
self-consistently and is a function of temperature.  
Note that $f_{\mathbf q}>0$ for all ${\mathbf q}$ 
if $n_{\mathrm ph}<1$.  
This implies that the MF theory is stable 
when there are not too many magnetophonons.  
The instability in the many-magnetophonon regime 
signals the onset of a ``depinning transition.''  
In fact, this instability is an artifact caused 
by the fact that our choice of the pinning potential $H_{p4}$ 
in Eq.\ (\ref{eq:Hp4}) has no lower bound.  
However, we believe there is indeed a depinning 
transition in real situations, and it is 
appropriate to regard $n_{\mathrm ph}^c=1$ 
as the critical number of 
magnetophonons at the transition in this model.  
As we go beyond the MF theory and include 
higher orders in $1/N$, $n_{\mathrm ph}$ 
decreases due to fluctuations in $Q_i$.  

\begin{figure}
\epsfxsize=3in
\centerline{ \epsffile{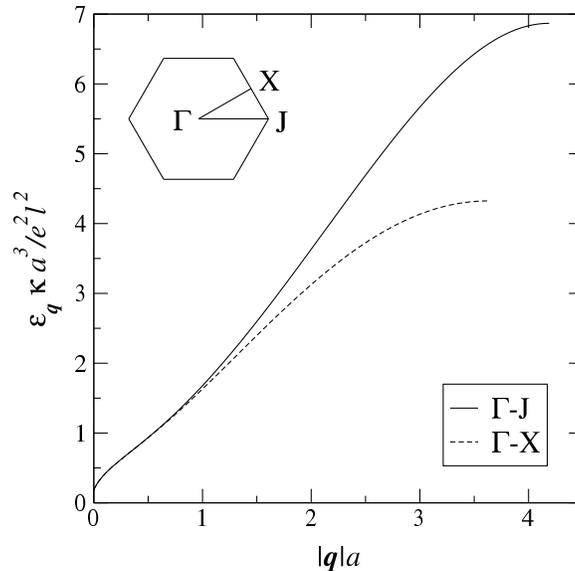} }
\caption{
The dispersion relation of a pinned triangular WC 
is plotted along two directions within the 
first Brillouin zone as specified in the inset.}
\label{fig:dispersion}
\end{figure}

Since the MF action $S_{\mathrm MF}$ is 
quadratic, it may be diagonalized using a 
Bogoliubov transformation.  It is more convenient 
to work in the Hamiltonian representation.  The 
Hamiltonian that reproduces $S_{\mathrm MF}$ is
\begin{eqnarray}
H_{\mathrm MF} = \sum_{\mathbf q} \left[ f_{\mathbf q} b_{\mathbf q}^\dagger b_{\mathbf q} - \frac{1}{2} \left( g_{\mathbf q} b_{-{\mathbf q}} b_{\mathbf q} + g_{\mathbf q}^* b_{\mathbf q}^\dagger b_{-{\mathbf q}}^\dagger \right) \right].
\end{eqnarray}
We define a new boson annihilation operator
\begin{equation}
\gamma_{\mathbf q} = u_{\mathbf q}b_{\mathbf q} + v_{\mathbf q}b_{-{\mathbf q}}^\dagger.
\end{equation}
In order to ensure the boson statistics, 
it has to satisfy
\begin{equation}
[\gamma_{\mathbf q},\gamma_{\mathbf q}^\dagger] = |u_{\mathbf q}|^2 - |v_{\mathbf q}|^2 = 1.
\end{equation}
By setting
\begin{eqnarray}
u_{\mathbf q} & = & e^{i\phi_{\mathbf q}} \sqrt{\frac{1}{2} \left( \frac{f_{\mathbf q}}{\varepsilon_{\mathbf q}} + 1 \right)}, \label{eq:u} \\
v_{\mathbf q} & = & \sqrt{\frac{1}{2} \left( \frac{f_{\mathbf q}}{\varepsilon_{\mathbf q}} - 1 \right)}, \label{eq:v}
\end{eqnarray}
we can diagonalize $H_{\mathrm MF}$ and write it as
\begin{equation}
H_{\mathrm MF} = \sum_{\mathbf q} \varepsilon_{\mathbf q} \gamma_{\mathbf q}^\dagger \gamma_{\mathbf q} + \text{const.},
\end{equation}
where $g_{\mathbf q}=|g_{\mathbf q}|e^{i\phi_{\mathbf q}}$.  
The energy eigenvalues are given by
\begin{equation}
\varepsilon_{\mathbf q} = \sqrt{f_{\mathbf q}^2 - |g_{\mathbf q}|^2}. \label{eq:energy}
\end{equation}
An example of the dispersion relation is shown in 
Fig.\ \ref{fig:dispersion}.  
Substituting $V=0$ in the above expressions, 
the dispersion relation without 
pinning, $\varepsilon_{\mathbf q} \propto |{\mathbf q}|^{3/2}$, 
is recovered. \cite{bonsall77,fukuyama75,maki83}  

Taylor-expanded in small $|{\mathbf q}|$, the 
parameters and the energy eigenvalue take the form
\begin{eqnarray}
f_{\mathbf q} & = & (1-\overline{Q})V + \frac{\nu e^2}{2\kappa} |{\mathbf q}| + {\mathcal O}[|{\mathbf q}|^2], \\
g_{\mathbf q} & = & \frac{\nu e^2}{2\kappa} \frac{(q_x+iq_y)^2}{|{\mathbf q}|} + {\mathcal O}[|{\mathbf q}|^2], \\
\varepsilon_{\mathbf q} & = & (1-\overline{Q})V + \frac{\nu e^2}{2\kappa} |{\mathbf q}| + {\mathcal O}[|{\mathbf q}|^2], \label{eq:smallq}
\end{eqnarray}
which are valid if
\begin{equation}
\frac{\nu e^2}{2\kappa} |{\mathbf q}| \ll (1-\overline{Q})V.
\end{equation}
One of the authors has obtained a similar dispersion 
relation at $T=0$. \cite{fertig99}  The main 
difference for this finite temperature model 
is that the zeroth order term contains the 
self-consistent MF value $\overline{Q}$.  
As pointed out in Ref.\ \ref{ref:fertig99}, 
the above linear dispersion is a direct consequence of 
the fact that the Coulomb interaction is long-ranged.  
With this dispersion relation, the magnetophonon 
density of states vanishes at 
the lower band edge, and thermal broadening will be 
substantially suppressed as will be shown below.  
This is not the case for a short-range interaction.  
For example, a screened interaction would have 
yielded a dispersion that is quadratic at small 
wave vectors.  In that case, the density of states jumps 
to a finite value at the band edge.  In a separate 
test, we have found that thermal broadening 
is indeed several orders of magnitude greater for 
a screened interaction than for an unscreened one.  

Since $f_{\mathbf q}$ and $\varepsilon_{\mathbf q}$ 
are functions of $\overline{Q}$, the saddle 
point solution for $\overline{Q}$ has to 
be determined self-consistently.  More specifically, 
it has to satisfy
\begin{eqnarray}
\overline{Q} & = & \frac{1}{N_s} \sum_{\mathbf q} \langle b_{l{\mathbf q}}^\dagger b_{l{\mathbf q}} \rangle \\
& = & \frac{1}{N_s} \sum_{\mathbf q} |u_{\mathbf q}|^2 \langle \gamma_{l{\mathbf q}}^\dagger \gamma_{l{\mathbf q}} \rangle + |v_{\mathbf q}|^2 \langle \gamma_{l,-{\mathbf q}} \gamma_{l,-{\mathbf q}}^\dagger \rangle \\
& = & \frac{1}{N_s} \sum_{\mathbf q} \left[ \left( n_B(\varepsilon_{\mathbf q}) + \frac{1}{2} \right) \frac{f_{\mathbf q}}{\varepsilon_{\mathbf q}} - \frac{1}{2} \right],
\end{eqnarray}
where $n_B(\varepsilon)=1/(e^{\beta\varepsilon}-1)$ 
is the Bose function.  

In the MF theory, the uniform field 
absorption spectrum will have a delta function 
peak at $\hbar\omega_p=\varepsilon_{\mathbf q=0}$.  
Since $\varepsilon_{\mathbf q=0}=V(1-\overline{Q})$ 
and $\overline{Q}$ increases with temperature, 
the thermal reduction of the peak frequency 
observed in experiments is explained qualitatively 
already at the MF level.  Physical interpretation is 
also simple.  At high temperatures, more 
magnetophonons are thermally created.  This 
implies that the average number of electrons that 
stay in the pits decreases, leading to a drop 
in the spatially averaged pinning potential.  

In order to explain thermal broadening, however, 
one has to go beyond the MF theory.  We will 
compute the first order corrections in 
the $1/N$ expansion in the next subsection.

\subsection{Order $1/N$ corrections}
\label{subsec:order1_N}

\begin{figure}
\epsfxsize=2.5in
\centerline{ \epsffile{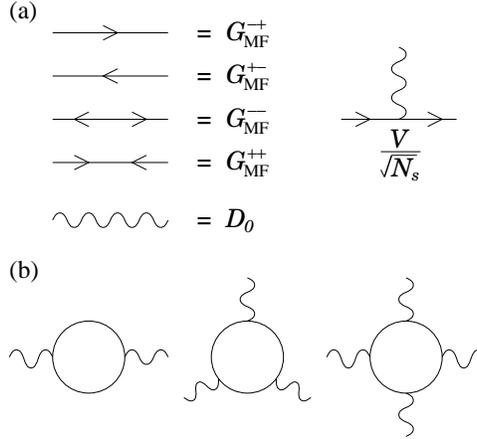} }
\caption{
(a) The Feynman diagrams are drawn for 
MF magnetophonon Green's functions $G_{\mathrm MF}$ 
(straight line), Hubbard-Stratonovich 
propagator $D_0$ (curly line), and the 
vertex $V/\sqrt{N_s}$.  The arrows on 
both sides of a vertex must be the same.  
Note that there are anomalous Green's functions 
that have two arrows pointing opposite 
directions, because the Hamiltonian does 
not conserve the magnetophonon number.  
(b) Examples of loops with vertices.  Each loop 
corresponds to a summand of the $n$-summation 
in Eq.\ (\ref{eq:SEff3}).  $n=2,3,$ and 4 respectively.  
A straight line without arrows 
denote a sum of all arrowed lines that are 
permitted by the ``direction-conservation'' 
rule at each vertex.  The first diagram 
represents the RPA propagator.  
Since there is no $n=1$ term in 
Eq.\ (\ref{eq:SEff3}), the Feynman rules 
do not allow a loop to have only one vertex.}
\label{fig:feynman}
\end{figure}

In this subsection, we will consider fluctuations 
of the HS field about its MF solution.  
The fluctuation fields are defined at each 
site as
\begin{equation}
r_i(\tau) = Q_i(\tau) - \overline{Q}
\end{equation}
and its Fourier-transformed field is
\begin{equation}
r_q = \frac{1}{\beta\sqrt{N_s}} \sum_i e^{-i{\mathbf q}\cdot{\mathbf R}_i} r_i(\tau).
\end{equation}
Since $r_i(\tau)$ is real, $r_{-q}=r_q^*$.  
In terms of $r_q$, 
Eq.\ (\ref{eq:SHS}) may be rewritten as
\begin{eqnarray}
S_{\mathrm HS} & = & -\frac{\beta}{2} \sum_l\sum_{qq'} \left[ \begin{array}{cc} z_{lq}^* & z_{l,-q} \end{array} \right] ({\mathsf G}_{\mathrm MF}^{-1} + {\mathsf U})_{qq'} \left[ \begin{array}{c} z_{lq'} \\ z_{l,-q'}^* \end{array} \right] \nonumber \\
& & + \frac{\beta NV}{2} \left( N_s\overline{Q}^2 - 2\sqrt{N_s}\overline{Q}r_{q=0} + \sum_q |r_q|^2 \right),
\end{eqnarray}
where
\begin{eqnarray}
({\mathsf G}_{\mathrm MF}^{-1})_{qq'} & = & \left[
  \begin{array}{cc}
  i\omega_n-f_{\mathbf q} & g_{\mathbf q}^* \\
  g_{\mathbf q} & -i\omega_n-f_{\mathbf q}
  \end{array}
\right] \delta_{qq'}, \\
{\mathsf U}_{qq'} & = & - \frac{1}{\sqrt{N_s}} \left[
  \begin{array}{cc}
  V r_{q-q'} & 0 \\
  0 & V r_{q-q'} \\
  \end{array}
\right].
\end{eqnarray}
${\mathsf G}_{\mathrm MF}$ is the MF magnetophonon 
Green's function matrix and ${\mathsf U}$ 
describes the coupling between a magnetophonon 
and the HS field $r$, or a vertex in the 
Feynman diagrams.  The ``bare'' 
HS propagator can be read off 
from $S_{\mathrm HS}$ and is given by
\begin{equation}
D_0 = -\frac{1}{NV}.
\end{equation}
Fig.\ \ref{fig:feynman}(a) shows the building 
blocks of the Feynman diagrams that will 
be frequently used below.  

Note that $D_0$ is proportional to $1/N$, 
so each curly line raises the order of an overall 
diagram by one in the $1/N$ expansion.  On the 
other hand, a magnetophonon loop 
[see Fig.\ \ref{fig:feynman}(b)] contributes 
order $N$ due to a summation over an internal 
species-index.  In order to find corrections to the HS 
propagator, it is convenient to integrate out the 
magnetophonon degree of freedom.  The 
effective action for the HS field $r_q$ is given by
\begin{eqnarray}
S_{\mathrm eff}[r] & = & -\ln \int {\mathcal D}^2z\, e^{-S_{\mathrm HS}} \\
& = & -\frac{N}{2} \text{tr}\,\ln \left[ \beta({\mathsf G}_{\mathrm MF}^{-1} + {\mathsf U}) \right] \nonumber \\
& & \quad + \frac{\beta NV}{2} \left( \sum_q |r_q|^2 - 2\overline{Q}\sqrt{N_s}r_{q=0} \right) \label{eq:SEff2} \\
& = & \frac{N}{2} \left[ \text{tr}\,\ln \frac{{\mathsf G}_{\mathrm MF}}{\beta} - \sum_{n=2}^\infty \frac{1}{n} \text{tr}\,({\mathsf G}_{\mathrm MF}{\mathsf U})^n \right] \nonumber \\
& & \quad + \frac{\beta NV}{2} \sum_q |r_q|^2. \label{eq:SEff3}
\end{eqnarray}
Going from Eq.\ (\ref{eq:SEff2}) to Eq.\ (\ref{eq:SEff3}), 
we have Taylor-series expanded the logarithmic function 
about $\beta{\mathsf G}_{\mathrm MF}^{-1}$.  
Unimportant constants were omitted at 
every step.  Note that the last term in Eq.\ (\ref{eq:SEff2}) 
exactly cancels out the $n=1$ term that would 
have been in Eq.\ (\ref{eq:SEff3}), 
because $\overline{Q}$ is chosen at 
a saddle point.  ${\mathsf G}_{\mathrm MF}$ and ${\mathsf U}$ 
are considered as entities 
in a product space of a $2\times 2$ matrix and 
a matrix labeled by frequency and wave-number indices $qq'$.  
Thus, a trace operator sums not only over the $2\times 2$ 
matrix indices, but also over $q$.  

Let us take a closer look at the last expression (\ref{eq:SEff3}).  
The first term is the MF free energy and the 
last term is the action of free independent HS fields $r_q$.  
Each of the remaining terms is represented by a loop 
with $n$ vertices as in Fig.\ \ref{fig:feynman}(b).  

\begin{figure}
\epsfxsize=3in
\centerline{ \epsffile{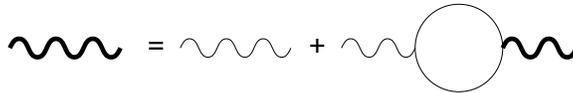} }
\caption{
Dyson equation for the RPA Hubbard-Stratonovich propagator.  
Thick wiggly lines denote self-consistent solution 
for the RPA propagator.}
\label{fig:RPA}
\end{figure}

Let us now search for the lowest order correction to 
the HS propagator.  There is no correction of 
the zeroth order or lower, and the only $1/N$ correction 
comes from the $n=2$ term of Eq.\ (\ref{eq:SEff3}).  
As can be seen in the first Feynman diagram in 
Fig.\ \ref{fig:feynman}(b), this 
term is nothing but the RPA correction 
to the HS propagator.  
We explicitly write it as
\begin{equation}
\frac{N}{4} \text{tr}\,{\mathsf G}_{\mathrm MF}{\mathsf U}{\mathsf G}_{\mathrm MF}{\mathsf U} = \frac{N\beta V^2}{2} \sum_q \varrho_{\mathrm MF}(q) |r_q|^2,
\end{equation}
where $\varrho_{\mathrm MF}(q)$ is the MF magnetophonon density-density 
correlation function defined as
\begin{eqnarray}
\varrho_{\mathrm MF}(q) & = & \frac{1}{2\beta N_s} \sum_{q'} \text{tr}\, {\mathsf G}_{\mathrm MF}(q'-q) {\mathsf G}_{\mathrm MF}(q') \label{eq:densityCorr} \\
& = & \int d\tau\, e^{i\omega_n\tau-i{\mathbf q}\cdot{\mathbf R}_i} \left\langle b_i^\dagger b_i (\tau) b_0^\dagger b_0 (0) \right\rangle_{\mathrm MF},
\end{eqnarray}
where $\langle\cdots\rangle_{\mathrm MF}$ denotes 
an average in the MF theory.  
Detailed calculations of $\varrho_{\mathrm MF}(q)$ will be given in 
Appendix A.  The RPA ``dressed'' propagator 
is defined through the Dyson equation
\begin{eqnarray}
D(q) & = & \frac{1}{D_0^{-1} + N V^2 \varrho_{\mathrm MF}(q)} \\
& = & -\frac{1}{NV ( 1 - V\varrho_{\mathrm MF}(q) )}.
\end{eqnarray}
Since both the bare propagator $D_0$ 
and the RPA correction are of order $1/N$, 
the above propagator is also of order $1/N$.  
A thick curly line will be used for 
the Feynman diagram of the dressed 
propagator as in Fig.\ \ref{fig:RPA}.  

\begin{figure}
\epsfxsize=2.5in
\centerline{ \epsffile{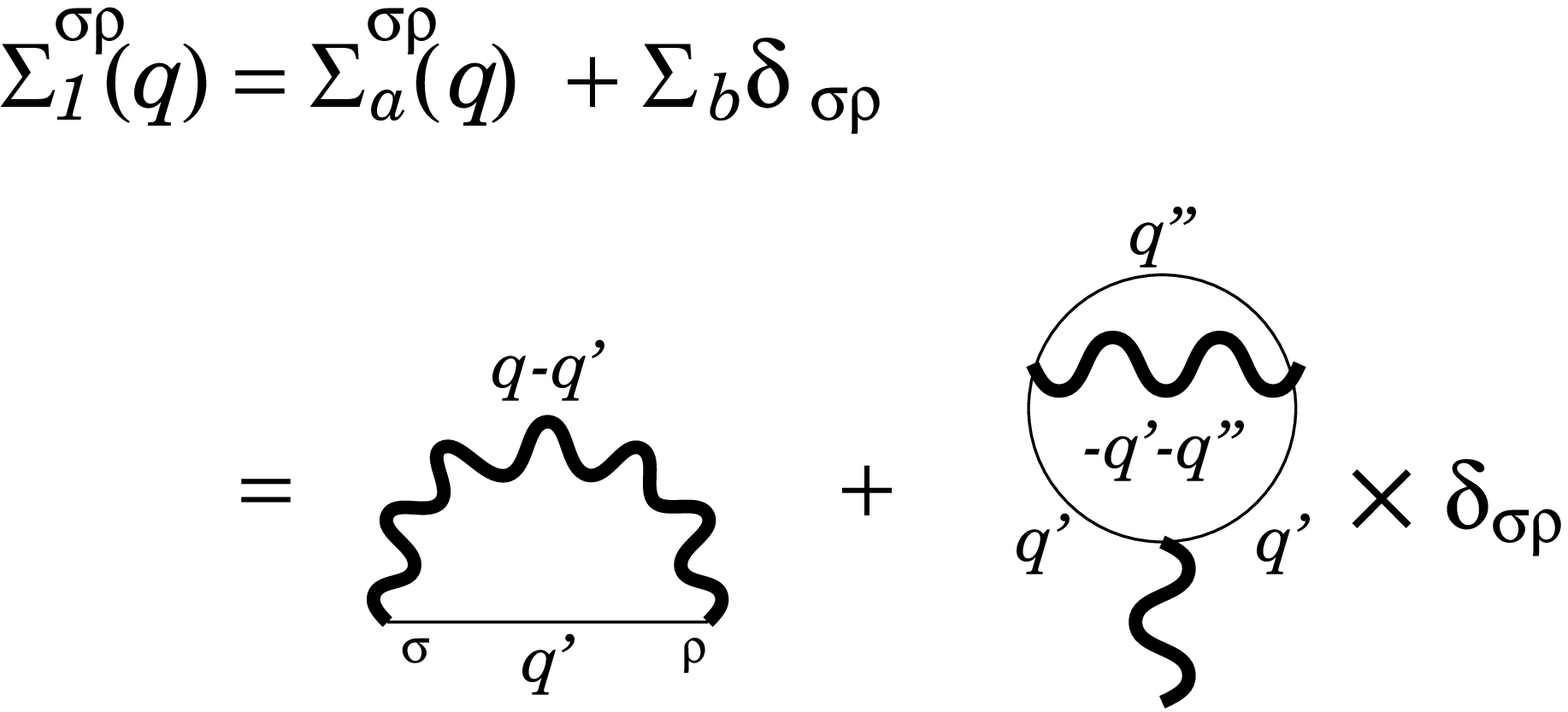} }
\caption{
Two self-energy diagrams of order $1/N$.  
The indices $\sigma$ and $\rho$ take either $+$ or $-$.  
They also determine the direction of arrows in the 
diagram of $\Sigma_a$.  A $-$ ($+$) 
sign means annihilation (creation) of 
a magnetophonon, so the corresponding 
arrow points towards 
(away from) the nearby vertex.}
\label{fig:selfEnergy}
\end{figure}

We now have all the ingredients to calculate 
the $1/N$ correction to the magnetophonon Green's 
function.  Keeping in mind that each HS propagator is 
of order $1/N$ and each magnetophonon loop is of order $N$, 
it is straightforward to count 
the order of any Feynman diagram.  There are 
two self-energy diagrams of order $1/N$ as 
shown in Fig.\ \ref{fig:selfEnergy}.  
They are written in terms of MF Green's 
functions $G_{\mathrm MF}^{\sigma\rho}$ 
and the HS propagator $D$ as
\begin{eqnarray}
\Sigma_a^{\sigma\rho}(q) & = & -\frac{V^2}{\beta N_s} \sum_{q'} D(q-q') G_{\mathrm MF}^{\sigma\rho}(q'), \label{eq:SigmaA} \\
\Sigma_b & = & \frac{NV^4}{\beta^2 N_s^2} D(0) \sum_{q'q''} D(-q'-q'') \nonumber \\
& & \ \times\biggl\{ G_{\mathrm MF}^{-+}(q'') \Bigl[ \bigl( G_{\mathrm MF}^{-+}(q') \bigr)^2 + G_{\mathrm MF}^{--}(q')G_{\mathrm MF}^{++}(q') \Bigr] \nonumber \\
& & \quad + \Bigl[ G_{\mathrm MF}^{--}(q'') G_{\mathrm MF}^{++}(q') G_{\mathrm MF}^{-+}(q') + \text{c.c.} \Bigr] \biggr\}. \label{eq:SigmaB}
\end{eqnarray}
These expressions can be partially evaluated 
analytically via frequency summations. \cite{frequencySum}  
More details may be found in Appendix B.  

The self-energy correction is incorporated into 
the Green's functions via the Dyson equation
\begin{equation}
{\mathsf G}(q) = \left[ {\mathsf G}_{\mathrm MF}^{-1}(q) - {\mathsf \Sigma}_a(q) - {\mathsf I}\Sigma_b \right]^{-1}, \label{eq:GDyson}
\end{equation}
where
\begin{equation}
{\mathsf \Sigma}_a = \left[ \begin{array}{cc} \Sigma_a^{-+} & \Sigma_a^{--} \\ \Sigma_a^{++} & \Sigma_a^{+-} \end{array} \right],
\qquad {\mathsf I} = \left[ \begin{array}{cc} 1 & 0 \\ 0 & 1 \end{array} \right].
\end{equation}

Combining Eq.\ (\ref{eq:GDyson}) with 
Eqs.\ (\ref{eq:sigma}), (\ref{eq:chiR}), and (\ref{eq:chi}), 
we can finally compute the ac conductivity to 
the order $1/N$.  Using the final expressions in Appendix B, 
we have numerically computed the self-energy 
diagrams and obtained the conductivity 
as a function of $T$.  The result is presented 
in the next section.

\section{Result}
\label{sec:result}

Before we present the result, 
it is useful to review some important 
energy scales.  For the purpose of comparison, 
we will use the data in Ref.\ \ref{ref:li98}.  
We find
\begin{eqnarray}
\frac{e^2}{\kappa l_0} & \gtrsim & 16\ {\mathrm meV}, \nonumber \\
\hbar\omega_c & \gtrsim & 3\ {\mathrm meV}, \nonumber \\
\frac{e^2}{\kappa a} & \sim & 2\ {\mathrm meV}, \nonumber \\
\frac{e^2l_0^2}{\kappa a^3} & \lesssim & 60\ \mu{\mathrm eV}, \nonumber \\
\hbar\omega_p & \sim & 5\ \mu{\mathrm eV}.
\end{eqnarray}
The temperature range for which the line 
width was clearly discernible 
was $3\ \mu$eV $\leq k_B T \lesssim 18\ \mu$eV.

For efficiency in numerical computations, 
we have made the following isotropic 
approximations
\begin{eqnarray}
f_{\mathbf q} & = & f(|{\mathbf q}|), \nonumber \\
g_{\mathbf q} & = & g(|{\mathbf q}|) \left(\frac{q_x+iq_y}{|{\mathbf q}|}\right)^2, \nonumber \\
\varepsilon_{\mathbf q} & = & \varepsilon(|{\mathbf q}|). \label{eq:isotropic}
\end{eqnarray}
We did, however, keep the shape of 
the first Brillouin zone to be hexagonal.  
According to our numerical tests, these 
approximations break down as the Brillouin 
zone boundaries are approached, 
i.e., when $|{\mathbf q}| \sim \pi/a$, 
but they are very accurate 
for $|{\mathbf q}|\lesssim 1/a$.  For example, 
when $V\sim\hbar\omega_p$, 
the energy $\varepsilon_{\mathbf q}$ 
is already an order of magnitude greater than $\hbar\omega_p$ at 
$|{\mathbf q}|=1/a$ (see Fig.\ \ref{fig:dispersion}), 
but the error $\Delta\varepsilon$ is still less than 4\%.   
Therefore, this approximation should 
be quantitatively reliable if $T$ is not 
too much greater than $\hbar\omega_p$.  

\begin{figure}
\epsfxsize=3in
\centerline{ \epsffile{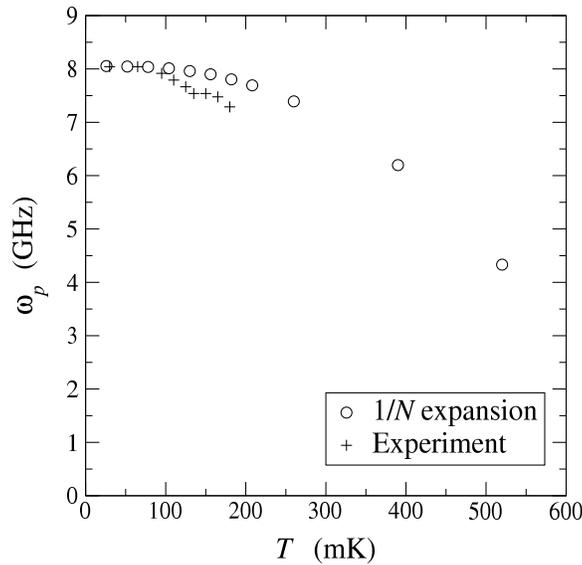} }
\caption{
Peak frequency 
as a function of temperature.  Our results 
are overestimated compared to the experimental 
data (Ref.\ \ref{ref:li98}), but there is a good 
qualitative agreement.}
\label{fig:peakPosition}
\end{figure}

In order to compare our result with the 
experimental data, we have chosen the bare 
pinning potential $V$ in such a way that the 
resulting peak frequency $\omega_p$ agrees 
with the experiment at the lowest measured 
temperature.  
The peak position data are shown in 
Fig.\ \ref{fig:peakPosition}, along with 
experimental data from Ref.\ \ref{ref:li98}.  
Although our results are systematically 
overestimated, the qualitative agreement is rather 
good.  

\begin{figure}
\epsfxsize=3in
\centerline{ \epsffile{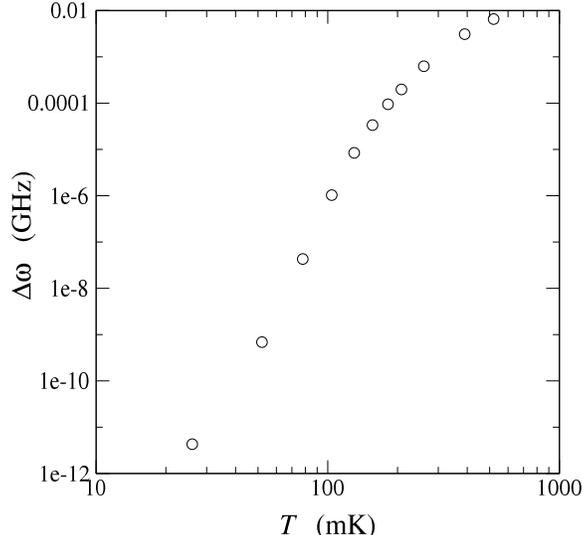} }
\caption{
The half-width-half-maximum line width 
is plotted against temperature in log-log 
scale.  The peaks appear 
to be extremely sharp especially in the low 
temperature regime.  However, the width 
increases very rapidly with temperature.}
\label{fig:peakWidth}
\end{figure}

The results for line width $\Delta\omega$ are surprisingly 
small (Fig.\ \ref{fig:peakWidth}).  Compared 
to those from experimental data, our results 
are several orders of magnitude smaller 
at the lowest $T$.  This is especially surprising 
because $T\sim\hbar\omega_p$.  Na\"{\i}vely, 
one would expect that the line width should 
be of the same order of magnitude as $T$, 
due to thermal broadening.  
Although much greater than our results, 
experiments also confirm that this is clearly 
not the case.  In Refs.\ \ref{ref:li97} and 
\ref{ref:li98}, $\Delta\omega$ is almost one 
order of magnitude smaller than $\omega_p$.  
In Ref.\ \ref{ref:hennigan98}, the quality 
factor $Q=\omega_p/\Delta\omega$ even exceeds 30.  
There are also other indications that the peaks 
may be actually even sharper. \cite{li99private}  

The unusually sharp peak may be rather well 
understood within our model.  As discussed 
in Sec.\ \ref{subsec:MFTheory}, due to the long-range 
Coulomb interaction, the dispersion 
starts out linearly in $|{\mathbf q}|$.  
The density of states thus increases linearly from 
zero.  Explicitly, the density of states per 
unit cell takes the form
\begin{equation}
{\mathcal N}(\varepsilon) = \left\{
  \begin{array}{ll} \displaystyle
  \frac{1}{2\pi} \left( \frac{2\kappa}{\nu e^2} \right)^2 \left[\varepsilon-(1-\overline{Q})V \right] & \text{if}\ \varepsilon>(1-\overline{Q})V, \\
  0 & \text{if}\ \varepsilon\leq (1-\overline{Q})V.
  \end{array}
\right.
\end{equation}
A rough estimate of the total number 
of ``accessible states'' per site at 
temperature $T$ would be
\begin{equation}
\sim a_c\int_0^T d\varepsilon\ {\mathcal N}(\varepsilon),
\end{equation}
where $a_c$ is the unit cell area.  Using the 
parameters in Ref.\ \ref{ref:li98}, we find 
that there are merely $2\times 10^{-4}$ 
accessible states per site at $T=180$\ mK, which 
is the highest temperature for which the 
line width was measured.  The physical interpretation 
of this may be given as the following.  At 
least in the semiclassical level, our model is 
essentially that of a crystal in which each electron 
is attached to a lattice site by a short-range 
binding potential.  
The overall uniform motion ($\mathbf q=0$) 
determines the peak frequency.  Broadening comes 
about from thermal and quantum fluctuations 
deforming the lattice and making the electrons 
move relative to one another.  However, the above analysis 
tells us that a WC in a strong magnetic field 
is so rigid that it is hardly deformed even at 
temperatures greater than the pinning potential.  

In reality, the external ac field has a finite 
wave length of order $\sim 30\ \mu$m $\sim 10^5a$ 
(Ref.\ \ref{ref:li97}).  
Accordingly, we have performed a similar analysis 
for the response at the finite wave vector, but 
we found no qualitative change.  

Defects in a WC such as interstitials, 
vacancies, dislocations, etc.\ may also affect 
broadening.  It is certainly true that their 
presence will soften the lattice and subsequently 
enhance magnetophonon excitations.  However, 
unless one is too close to the melting temperature 
of the WC, which is not the case in the experiments, 
their effects may be mostly taken into account 
through renormalization of the lattice stiffness, or 
dynamic matrix in Eq.\ (\ref{eq:dynamicMatrix}).  
This should not affect our results qualitatively.  

There may be other sources of broadening, too.  
Other low energy modes, which are not included in 
this work, may be involved in 
the broadening.  One possible candidate is the 
edge states of the WC, \cite{fertig93} the analysis 
of which will be given elsewhere.  Another possible 
source of broadening is extrinsic low energy 
modes {\em outside} the WC.  When an 
experiment is performed, there 
are many external degrees of freedom that 
may be coupled to the sample.  If some of them 
are hard to remove and thus are left undetected to 
influence the data, the observed peak will 
be certainly broader than theoretically predicted.  

It has also been 
suggested that higher Landau level mixing might 
be important. \cite{fogler99}  It is certainly 
true that there can be second or higher order scattering 
processes that use higher Landau levels as 
virtual states.  However, these processes 
should be suppressed as a power $(\omega_p/\omega_c)^a$ 
with $a\geq 1$.  Since $\omega_p/\omega_c\sim 10^{-3}$, 
$\Delta\omega$ as a result of such processes would 
be still too small to explain currently available 
experimental data.

\section{Conclusion}
\label{sec:conclusion}

In this work, we studied the thermal broadening 
of the electromagnetic absorption resonance of 
a magnetically induced WC.  In the strong 
magnetic field limit, the low-energy collective modes 
that are coupled to the spatially uniform ac electric field 
are magnetophonons.  Assuming that 
electrons are closely bound to lattice sites, 
magnetophonon creation and annihilation operators 
may be constructed out of 
displacements ${\mathbf u}_i={\mathbf r}_i-{\mathbf R}_i$. 

We showed that 
the Hamiltonian could be divided into harmonic and 
anharmonic parts in terms of magnetophonon 
operators.  The harmonic part, when taken 
alone, describes independent magnetophonons 
and produces a delta function peak in the spectrum.  
The anharmonic part introduces magnetophonon interactions.  
It will not only renormalize the peak 
position, but also broaden it by mixing different 
magnetophonon modes.  Anharmonicity in our model 
comes from two sources: the Coulomb interaction and 
the pinning potential.  Since the Coulomb interaction 
depends only on the relative coordinates, it is 
completely decoupled from the center-of-mass 
degrees of freedom.  This leads to the 
derivation of a generalized Kohn's theorem 
that asserts that the Coulomb interaction 
cannot shift or broaden an ac 
conductivity resonance peak even in the presence 
of a uniform harmonic pinning potential.  

Analysis of the magnetophonon interactions 
in the pinning potential was 
performed using a $1/N$ expansion technique.  
This technique provides a systematic way 
of summing up important diagrams and it 
captures important pieces of physics in the 
low order solutions.  For example, we found that 
the zeroth order MF solution could account for 
the decreasing peak frequency $\omega_p$ as 
a function of temperature.  

Thermal broadening 
appears in the $1/N$ self-energy corrections.  
However, the line 
width $\Delta\omega$ is found to be many orders 
of magnitude smaller than $\omega_p$ and $T$.  
The reason for this lies in the magnetophonon 
dispersion that is linear and steep in the low energy 
limit.  This translates into a small density of 
states for collective excitations.  Consequently 
the broadening is substantially suppressed.  

Our result of peak frequency agrees qualitatively with recent 
experiments. \cite{li97,li98}  The line width 
result is also qualitatively consistent in that it is 
much less than $\omega_p$ and $T$.  Quantitatively, 
our result of $\Delta\omega$ appears to be much smaller 
than published results, although there is continuing 
experimental work on sorting out just how narrow 
the intrinsic line width really is. \cite{li99private}  
Other possible sources of broadening, such as edge 
states and extrinsic low energy modes, may also 
be responsible for observed line width.

\section*{Acknowledgment}
We are indebted to Lloyd Engel, Chi-Chun Li, Dan Tsui, 
and Jongsoo Yoon for helpful discussions regarding 
their experiments.  Carsten Timm and Steve Girvin 
are gratefully acknowledged for useful discussions 
about the $1/N$ expansion technique.  
We also thank Institute for Theoretical Physics, UC 
Santa Barbara, where 
part of this work was done.  This work was supported 
by NSF Grant DMR98-70681 and PHY94-07194 and the 
Research Corporation.

\appendix

\section{Density-density correlation $\varrho_{\mathrm MF}(\lowercase{q})$}

\begin{figure}
\epsfxsize=2.5in
\centerline{ \epsffile{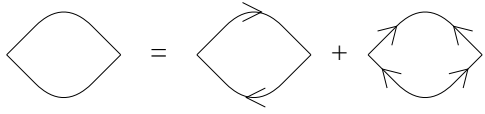} }
\caption{
Two arrowed diagrams for the density-density correlation 
function $\varrho_{\mathrm MF}(q)$.}
\label{fig:densityCorrelation}
\end{figure}

In this Appendix, 
we will present detailed calculations of the 
density-density correlation function $\varrho_{\mathrm MF}(q)$.  
Eq.\ (\ref{eq:densityCorr}) can be written as
\begin{eqnarray}
\varrho_{\mathrm MF}(q) & = & \frac{1}{2\beta N_s} \sum_{q'} \bigl[ G_{\mathrm MF}^{-+}(q'-q) G_{\mathrm MF}^{-+}(q') + G_{\mathrm MF}^{--}(q'-q) G_{\mathrm MF}^{++}(q') + G_{\mathrm MF}^{++}(q'-q) G_{\mathrm MF}^{--}(q') + G_{\mathrm MF}^{+-}(q'-q) G_{\mathrm MF}^{+-}(q') \bigr] \nonumber \\
& = & \frac{1}{\beta N_s} \sum_{q'} \bigl[ G_{\mathrm MF}^{-+}(q'-q) G_{\mathrm MF}^{-+}(q') + G_{\mathrm MF}^{--}(q'-q) G_{\mathrm MF}^{++}(q'), \bigr]
\end{eqnarray}
where we have used
\begin{equation}
G_{\mathrm MF}^{\sigma\rho}(-q) = G_{\mathrm MF}^{\rho\sigma}(q)
\end{equation}
to get the second line.  Diagrammatically, 
the last two terms are represented by two 
arrowed diagrams as in Fig.\ \ref{fig:densityCorrelation}.  

Using the Green's functions in Eq.\ (\ref{eq:MFGreensFunc}), 
we may rewrite the above equation more explicitly as
\begin{eqnarray}
\varrho_{\mathrm MF}(q) & = & \frac{1}{\beta N_s} \sum_{q'} \frac{2(i\omega_n'-i\omega_n)i\omega_n' + 2f_{\mathbf q'-q}f_{\mathbf q'} + g^*_{\mathbf q'-q}g_{\mathbf q'} + g_{\mathbf q'-q}g^*_{\mathbf q'}}{2[(i\omega_n'-i\omega_n)^2-\varepsilon_{\mathbf q'-q}^2][(i\omega_n')^2-\varepsilon_{\mathbf q'}^2]} \\
& = & \frac{1}{2N_s} \sum_{\mathbf q'} \left[ (n_{\mathbf q'-q}-n_{\mathbf q'}) \frac{(\varepsilon_{\mathbf q'-q}-\varepsilon_{\mathbf q'})(1+\xi_{\mathbf q'-q,q'})}{(i\omega_n)^2-(\varepsilon_{\mathbf q'-q}-\varepsilon_{\mathbf q'})^2} + (n_{\mathbf q'-q}+n_{\mathbf q'}+1) \frac{(\varepsilon_{\mathbf q'-q}+\varepsilon_{\mathbf q'})(1-\xi_{\mathbf q'-q,q'})}{(i\omega_n)^2-(\varepsilon_{\mathbf q'-q}+\varepsilon_{\mathbf q'})^2} \right],
\end{eqnarray}
where $n_{\mathbf q'}$ is a shorthand notation 
of the Bose function $n_B(\varepsilon_{\mathbf q'})$.  
In the second line, we have used the usual 
frequency summation technique. \cite{frequencySum}  

$\varrho(q)$ has many simple poles in 
the complex $\omega$ plane on the real axis.  
In the thermodynamic limit, 
they become branch cuts.  Therefore, 
summations involving $\varrho(q)$ must be 
performed with care, as will be demonstrated 
in the calculations of self energy corrections 
in the next Appendix.  
Note that $\varrho(q)$ is analytic 
and real at $\omega=0$, so that the branch cuts are divided into 
two: one on the positive and the other on the negative 
axis.  Both branch cuts are bounded 
because $\varepsilon_{\mathbf q'-q}\pm\varepsilon_{\mathbf q'}$ 
is finite.  

Eventually, we will analytically 
continue $i\omega_n\rightarrow\omega +i0^+$ 
to get a retarded correlation function.  
Since $\varrho(-q)=\varrho(q)$ 
and $\varrho(q)$ is analytic except on the branch cuts, 
it has the following properties:
\begin{eqnarray}
{\mathrm Re}\,\varrho(-\omega+i0^+,{\mathbf q}) & = & {\mathrm Re}\,\varrho(\omega+i0^+,{\mathbf q}), \nonumber \\
{\mathrm Im}\,\varrho(-\omega+i0^+,{\mathbf q}) & = & -{\mathrm Im}\,\varrho(\omega+i0^+,{\mathbf q}), \label{eq:rhoAnalyticity}
\end{eqnarray}
which will become useful in the next Appendix.  
Finally, $\varrho(\omega +i0^+,{\mathbf q})$ 
was numerically computed using an isotropic 
approximation as explained in 
Sec.\ \ref{sec:result}.

\section{$1/N$ Self energy corrections $\Sigma_a^{\sigma\rho}$ and $\Sigma_b$}

In this Appendix, we will calculate the $1/N$ 
self energy corrections to the magnetophonon 
Green's functions as in Fig.\ \ref{fig:selfEnergy}.  
Since the bare HS propagator $D_0$ is a constant, 
it is convenient to separate it from the rest 
of the RPA dressed propagator $D$.  The remaining part
\begin{eqnarray}
\delta D({\mathbf q}) & \equiv & D({\mathbf q}) - D_0 \\
& = & \frac{\varrho_0({\mathbf q})}{N[V\varrho_0({\mathbf q})-1]}
\end{eqnarray}
vanishes as $\omega^{-2}$ in the large $\omega$ limit, 
so all following frequency sums that contain $\delta D({\mathbf q})$ 
converge without introducing 
cumbersome convergence factors such as $e^{\pm i\omega 0^+}$.  

The first self-energy correction $\Sigma_a^{\sigma\rho}(q)$ 
is given in Eq.\ (\ref{eq:SigmaA}).  
The contribution of the bare propagator $D_0$ is 
simply
\begin{equation}
\Sigma^{\sigma\rho}_{a0} = -\frac{V^2}{\beta N_s} D_0 \sum_{q'} G^{\sigma\rho}_{\mathrm MF}(q').
\end{equation}
The summation of MF Green's function 
is defined with an appropriate convergence 
factor that ensures normal-ordering.  
It is then straightforward to show
\begin{eqnarray}
\sum_{q'} G^{-+}_{\mathrm MF}(q') & = & -\beta \sum_{\mathbf q'} \langle b^\dagger_{\mathbf q'} b_{\mathbf q'} \rangle = -\beta N_s n_{\mathrm ph} \label{eq:sumG1}, \\
\sum_{q'} G^{+-}_{\mathrm MF}(q') & = & -\beta N_s n_{\mathrm ph}, \\
\sum_{q'} G^{--}_{\mathrm MF}(q') & = & -\beta \sum_{\mathbf q'} \langle b_{\mathbf q'} b_{\mathbf q'} \rangle \equiv -\beta N_s n_{\mathrm an}, \\
\sum_{q'} G^{--}_{\mathrm MF}(q') & = & -\beta \sum_{\mathbf q'} \langle b^\dagger_{\mathbf q'} b^\dagger_{\mathbf q'} \rangle = -\beta N_s n_{\mathrm an}^*, \label{eq:sumG4}
\end{eqnarray}
where $n_{\mathrm an}$ is the 
anomalous magnetophonon number per site.  
Due to the six-fold symmetry of the 
lattice, however, $n_{\mathrm an}=0$.  
We therefore get
\begin{eqnarray}
\Sigma^{-+}_{a0} & = & \Sigma^{+-}_{a0} = -\frac{n_{\mathrm ph}V}{N}, \nonumber \\
\Sigma^{--}_{a0} & = & \Sigma^{++}_{a0} = 0. \label{eq:SigmaA0}
\end{eqnarray}

\begin{figure}
\epsfxsize=3in
\centerline{ \epsffile{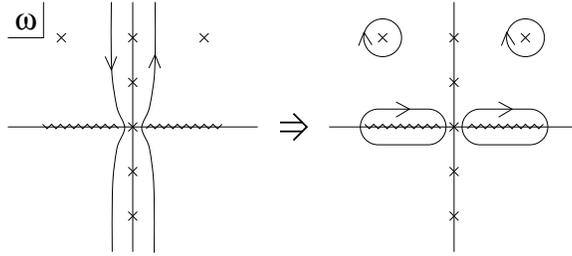} }
\caption{
Contour is deformed in the complex $\omega$.  
Simple poles are denoted by $\times$ and branch cuts 
by zigzag lines.  The separation between two branch 
cuts is exaggerated in order to show that the 
pole at the origin is treated separately.  In 
fact, the branch cuts asymptotically touch the origin.}
\label{fig:contour}
\end{figure}

Now let us turn our attention to the remaining 
part.  It needs to be computed with more care, 
since $\delta D(q')$ has branch cuts on the 
real $\omega$ axis, that arise from 
the density-density correlation function $\varrho(q')$.  
Using the Green's functions in Eq.\ (\ref{eq:MFGreensFunc}), 
we may write
\begin{eqnarray}
\delta\Sigma^{-+}_a(q) & = & -\frac{V^2}{N\beta N_s} \sum_{q'} \delta D(q-q') G^{-+}_{\mathrm MF}(q') \\
& = & -\frac{V^2}{N\beta N_s} \sum_{q'} \delta D(q-q') \left[ \frac{|u_{\mathbf q'}|^2}{i\omega_n'-\varepsilon_{\mathbf q'}} - \frac{|v_{\mathbf q'}|^2}{i\omega_n'+\varepsilon_{\mathbf q'}} \right],
\end{eqnarray}
where $u_{\mathbf q'}$ and $v_{\mathbf q'}$ are defined 
in Eqs.\ (\ref{eq:u}) and (\ref{eq:v}).  As 
in a standard frequency summation technique, \cite{frequencySum} 
we may replace the sum over $i\omega_n'$ by 
a contour integral on the complex $\omega'$ 
plane with simple poles on the imaginary axis.  
After the contour is deformed as in Fig.\ \ref{fig:contour}, 
the new expression takes the form
\begin{eqnarray}
\delta\Sigma^{-+}_a(q) & = & \frac{V^2}{NN_s} \sum_{\mathbf q'} \left[ \delta D(i\omega_n-\varepsilon_{\mathbf q'},{\mathbf q-q'})|u_{\mathbf q'}|^2 + \delta D(i\omega_n+\varepsilon_{\mathbf q'},{\mathbf q-q'})|v_{\mathbf q'}|^2 \right] n_B(\varepsilon_{\mathbf q'}) \nonumber \\
& & - \frac{V^2}{\pi NN_s} \sum_{\mathbf q'} \int_{\omega'\neq 0} d\omega' [{\mathrm Im}\,\delta D(\omega'+i0^+,{\mathbf q-q'})] \left[ \frac{|u_{\mathbf q'}|^2 n_B(\omega')}{i\omega_n+\omega'-\varepsilon_{\mathbf q'}} + \frac{|v_{\mathbf q'}|^2 n_B(-\omega')}{i\omega_n+\omega'+\varepsilon_{\mathbf q'}} \right], \label{eq:contourIntegral}
\end{eqnarray}
where $\omega'=0$ is excluded from 
the domain of the integral
\begin{equation}
\int_{\omega'\neq 0} = \int_{-\infty}^{0^-} + \int_{0^+}^{\infty}.
\end{equation}
Although $n_B(\omega')$ 
diverges as $\omega'\rightarrow 0$, 
the above integral is well defined, 
because ${\mathrm Im}\,\delta D(\omega'+i0^+,{\mathbf q-q'})$ 
vanishes as $\omega'=0$ is approached.  
Finally, analytic continuation is performed 
to give
\begin{eqnarray}
\delta\Sigma^{-+}_a(\omega+i0^+,{\mathbf q}) & = & \frac{V^2}{NN_s} \sum_{\mathbf q'} \left[ \delta D(\omega-\varepsilon_{\mathbf q'}+i0^+,{\mathbf q-q'})|u_{\mathbf q'}|^2 + \delta D(\omega+\varepsilon_{\mathbf q'}+i0^+,{\mathbf q-q'})|v_{\mathbf q'}|^2 \right] n_B(\varepsilon_{\mathbf q'}) \nonumber \\
& & - \frac{V^2}{\pi NN_s} \sum_{\mathbf q'} P\!\!\int d\omega' {\mathrm Im}\,\delta D(\omega'+i0^+,{\mathbf q-q'}) \left[ \frac{|u_{\mathbf q'}|^2 n_B(\omega')}{\omega+\omega'-\varepsilon_{\mathbf q'}} + \frac{|v_{\mathbf q'}|^2 n_B(-\omega')}{\omega+\omega'+\varepsilon_{\mathbf q'}} \right] \nonumber \\
& & -i\frac{V^2}{NN_s} \sum_{\mathbf q'} \Bigl[ {\mathrm Im}\,\delta D(\omega-\varepsilon_{\mathbf q'}+i0^+,{\mathbf q-q'}) |u_{\mathbf q'}|^2 n_B(\varepsilon_{\mathbf q'}-\omega) \nonumber \\
& & \qquad\qquad\qquad + {\mathrm Im}\,\delta D(\omega+\varepsilon_{\mathbf q'}+i0^+,{\mathbf q-q'}) |v_{\mathbf q'}|^2 n_B(\varepsilon_{\mathbf q'}+\omega) \Bigr], \label{eq:dSigmaA}
\end{eqnarray}
where $P\!\!\int$ denotes a Cauchy principal 
integral.  Eq.\ (\ref{eq:rhoAnalyticity}) 
was used in the last line.  The above quantity may 
be computed using the numerical solution 
of $\varrho(\omega+i0^+,{\mathbf q})$.  The 
other $\delta\Sigma_a$'s can be calculated 
in a similar way.  The final form 
of $\delta\Sigma^{+-}_a$ is the same 
the above equation, except that $|u_{\mathbf q'}|^2$ 
and $|v_{\mathbf q'}|^2$ are switched.  
$\delta\Sigma^{--}_a$ ($\delta\Sigma^{++}_a$) 
is obtained by replacing both $|u_{\mathbf q'}|^2$ 
and $|v_{\mathbf q'}|^2$ by $u_{\mathbf q'}^*v_{\mathbf q'}$ 
($u_{\mathbf q'}v_{\mathbf q'}^*$).  In the isotropic 
approximation, the angular sum of
\begin{equation}
u_{\mathbf q'}^*v_{\mathbf q'} = \frac{g_{\mathbf q'}^*}{2\varepsilon_{\mathbf q'}} \propto q_x-iq_y
\end{equation}
vanishes, so
\begin{equation}
\delta\Sigma^{--}_a = \delta\Sigma^{++}_a = 0.
\end{equation}

The other $1/N$ correction, $\Sigma_b$, may be calculated 
in a similar manner.  First, we compute the 
contribution from the bare HS propagator.  
Substituting $D_0$ for $D(-q'-q'')$, 
Eq.\ (\ref{eq:SigmaB}) becomes
\begin{equation}
\Sigma_{b0} = -\frac{V^3}{\beta^2 N_s^2} D(0) \sum_{q'q''} \biggl\{ G_{\mathrm MF}^{-+}(q'') \Bigl[ \bigl( G_{\mathrm MF}^{-+}(q') \bigr)^2 + G_{\mathrm MF}^{--}(q)G_{\mathrm MF}^{++}(q') \Bigr] + \Bigl[ G_{\mathrm MF}^{--}(q'') G_{\mathrm MF}^{++}(q') G_{\mathrm MF}^{-+}(q') + \text{c.c.} \Bigr] \biggr\}.
\end{equation}
Using Eqs.\ (\ref{eq:sumG1}) through (\ref{eq:sumG4}) along with
\begin{eqnarray}
\sum_{q'} \bigl( G_{\mathrm MF}^{-+}(q') \bigr)^2 & = & \sum_{q'} \left( \frac{|u_{\mathbf q'}|^2}{i\omega_n-\varepsilon_{\mathbf q'}} - \frac{|v_{\mathbf q'}|^2}{i\omega_n+\varepsilon_{\mathbf q'}} \right)^2 \nonumber \\
& = & -\beta \sum_{\mathbf q'} \left[ \left( |u_{\mathbf q'}|^4+|v_{\mathbf q'}|^4 \right) n_B'(\varepsilon_{\mathbf q'}) - 2|u_{\mathbf q'}|^2|v_{\mathbf q'}|^2 \frac{2n_B(\varepsilon_{\mathbf q'})+1}{2\varepsilon_{\mathbf q'}} \right], \\
\sum_{q'} G_{\mathrm MF}^{--}(q') G_{\mathrm MF}^{++}(q') & = & \sum_{q'} |u_{\mathbf q'}|^2 |v_{\mathbf q'}|^2 \left( \frac{1}{i\omega_n-\varepsilon_{\mathbf q'}} - \frac{1}{i\omega_n+\varepsilon_{\mathbf q'}} \right)^2 \nonumber \\
& = & -\beta \sum_{\mathbf q'} |u_{\mathbf q'}|^2 |v_{\mathbf q'}|^2 \left[ n_B'(\varepsilon_{\mathbf q'}) - \frac{2n_B(\varepsilon_{\mathbf q'})+1}{2\varepsilon_{\mathbf q'}} \right],
\end{eqnarray}
we get
\begin{equation}
\Sigma_{b0} = \frac{V^2 n_{\mathrm ph}}{[V\varrho(0)-1]NN_s} \sum_{\mathbf q'} \left\{ \frac{|g_{\mathbf q'}|^2}{\varepsilon_{\mathbf q'}^2} \frac{2n_B(\varepsilon_{\mathbf q'})+1}{2\varepsilon_{\mathbf q'}} - \frac{f_{\mathbf q'}^2}{\varepsilon_{\mathbf q'}^2} n_B'(\varepsilon_{\mathbf q'}) \right\}. \label{eq:SigmaB0}
\end{equation}
The derivative of the Bose function 
satisfies $n_B'(\varepsilon_{\mathbf q'})=-\beta n_B(\varepsilon_{\mathbf q'})[n_B(\varepsilon_{\mathbf q'})+1]$.  

For the remaining part, we 
substitute $\delta D(-q'-q'')$ for $D(-q'-q'')$.  
We get
\begin{eqnarray}
\delta\Sigma_b & = & \frac{NV^4}{\beta^2N_s^2} D(0) \sum_{q'q''} \frac{\delta D(-q'-q'')}{i\omega_n''-\varepsilon''} \Biggl[ \frac{|u'u''^*+v'v''^*|^2 \left(|u'|^2+|v'|^2\right)}{(i\omega_n'+\varepsilon')^2} + \frac{|u'v''+v'u''|^2 \left(|u'|^2+|v'|^2\right)}{(i\omega_n'-\varepsilon')^2} \nonumber \\
& & \qquad\qquad\qquad\qquad\qquad\qquad\qquad\qquad - \frac{2(u'u''^*+v'v''^*)(u'v''+v'u'')u'^*v'^* + \text{c.c.}}{(i\omega_n')^2-\varepsilon'^2} \Biggr].
\end{eqnarray}
Shorthand notations are defined as $u'\equiv u_{\mathbf q'}$, 
$u''\equiv u_{\mathbf q''}$, etc.
In order to avoid the branch cuts in $\delta D$, 
we make a change of 
variables $i\omega_n'\rightarrow -i\omega_n'-i\omega_n''$ 
and perform the frequency sum over $i\omega_n''$.  
The result takes the form
\begin{eqnarray}
\delta\Sigma_b & = & -\frac{NV^4}{\beta N_s^2} D(0) \sum_{q'{\mathbf q''}} \delta D(i\omega_n',{\mathbf -q'-q''}) \Biggl\{ |u'u''^*+v'v''^*|^2 \left(|u'|^2+|v'|^2\right) \left[ \frac{n_B(\varepsilon'')-n_B(\varepsilon')}{(i\omega_n'-\varepsilon'+\varepsilon'')^2} - \frac{n_B'(\varepsilon')}{i\omega_n'-\varepsilon'+\varepsilon''} \right] \nonumber \\
& & \qquad\qquad\qquad + |u'v''+v'u''|^2 \left(|u'|^2+|v'|^2\right) \left[ \frac{n_B(\varepsilon'')+n_B(\varepsilon')+1}{(i\omega_n'+\varepsilon'+\varepsilon'')^2} - \frac{n_B'(\varepsilon')}{i\omega_n'+\varepsilon'+\varepsilon''} \right] \nonumber \\
& & \qquad\qquad\qquad - \frac{(u'u''^*+v'v''^*)(u'v''+v'u'')u'^*v'^* + \text{c.c.}}{\varepsilon'} \left[ \frac{n_B(\varepsilon'')-n_B(\varepsilon')}{i\omega_n'-\varepsilon'+\varepsilon''} - \frac{n_B(\varepsilon'')+n_B(\varepsilon')+1}{i\omega_n'+\varepsilon'+\varepsilon''} \right] \Biggr\}. \label{eq:deltaSigmaB}
\end{eqnarray}
The last frequency sum $\sum_{i\omega_n'}$ is 
performed using a similar contour deformation 
technique as in Fig.\ \ref{fig:contour}.  Using
\begin{eqnarray}
\sum_{i\omega_n'} \frac{\delta D(i\omega_n')}{i\omega_n'-\varepsilon} & = & \frac{\beta}{\pi} {\mathrm Im}\int_{\omega'\neq 0} d\omega'\,\frac{\delta D(\omega'+i0^+)n_B(\omega')}{\omega'-\varepsilon+i0^+} \\
& = & \frac{\beta}{\pi} P\!\!\int d\omega'\,\frac{{\mathrm Im}\,\delta D(\omega'+i0^+)n_B(\omega')}{\omega'-\varepsilon} - \beta{\mathrm Re}\,\delta D(\varepsilon+i0^+)n_B(\varepsilon), \\
\sum_{i\omega_n'} \frac{\delta D(i\omega_n')}{(i\omega_n'-\varepsilon)^2} & = & \frac{\beta}{\pi} {\mathrm Im}\int_{\omega'\neq 0} d\omega'\,\left[ \frac{\partial}{\partial\omega'} \delta D(\omega'+i0^+)n_B(\omega') \right] \frac{1}{\omega'-\varepsilon+i0^+},
\end{eqnarray}
Eq.\ (\ref{eq:deltaSigmaB}) becomes
\begin{eqnarray}
\delta\Sigma_b & = & -\frac{V^3}{[V\varrho(0)-1]N_s^2} \sum_{\mathbf q'q''}{\mathrm Im}\int_{\omega'\neq 0} d\omega'\, \Biggl\{ \delta D(\omega',{\mathbf -q'-q''})n_B(\omega') \left( \frac{c_1}{\omega'-\varepsilon'+\varepsilon''+i0^+} + \frac{c_2}{\omega'+\varepsilon'+\varepsilon''+i0^+} \right) \nonumber \\
& & \qquad\qquad\qquad\qquad\qquad + \left[ \frac{\partial}{\partial\omega'} \delta D(\omega',{\mathbf -q'-q''})n_B(\omega') \right] \left( \frac{c_3}{\omega'-\varepsilon'+\varepsilon''+i0^+} + \frac{c_4}{\omega'+\varepsilon'+\varepsilon''+i0^+} \right), \label{eq:dSigmaB}
\end{eqnarray}
where
\begin{eqnarray}
c_1 & = & -|u'u''^*+v'v''^*|^2 \left(|u'|^2+|v'|^2\right) n_B'(\varepsilon') + \frac{(u'u''^*+v'v''^*)(u'v''+v'u'')u'^*v'^* + \text{c.c.}}{\varepsilon'} [n_B(\varepsilon')-n_B(\varepsilon'')], \\
c_2 & = & -|u'v''+v'u''|^2 \left(|u'|^2+|v'|^2\right) n_B'(\varepsilon') + \frac{(u'u''^*+v'v''^*)(u'v''+v'u'')u'^*v'^* + \text{c.c.}}{\varepsilon'} [n_B(\varepsilon')+n_B(\varepsilon'')+1], \\
c_3 & = & -|u'u''^*+v'v''^*|^2 \left(|u'|^2+|v'|^2\right) [n_B(\varepsilon')-n_B(\varepsilon'')], \\
c_4 & = & |u'v''+v'u''|^2 \left(|u'|^2+|v'|^2\right) [n_B(\varepsilon')+n_B(\varepsilon'')+1].
\end{eqnarray}

Finally, combining Eqs.\ (\ref{eq:SigmaA0}), (\ref{eq:dSigmaA}), 
(\ref{eq:SigmaB0}), and (\ref{eq:dSigmaB}), 
the order $1/N$ self-energy correction is
given by
\begin{eqnarray}
\Sigma_1^{\sigma\rho} = \Sigma_{a0}^{\sigma\rho} + \delta\Sigma_a^{\sigma\rho} + (\Sigma_{b0} + \delta\Sigma_b)\delta_{\sigma\rho}.
\end{eqnarray}

\end{document}